\documentclass[times,12pt]{article}

\usepackage{mdwlist}
\usepackage[utf8]{inputenc}

\usepackage{times,natbib}
\usepackage{epsfig}
\usepackage{graphicx}
\usepackage{amsmath, amsthm}
\usepackage{amssymb}
\usepackage{setspace}
\usepackage[margin=1in]{geometry}
\RequirePackage[colorlinks,citecolor=blue,urlcolor=blue]{hyperref}
\usepackage{natbib}
\usepackage{hyperref}
\usepackage{booktabs}
\usepackage{lineno}
\usepackage{multirow}

\newcommand{\bfg}{{\bf g}}

\def\bSig\mathbf{\Sigma}

\newcommand{\pkg}[1]{{\fontseries{b}\selectfont #1}}
\let\proglang=\textsf

\begin{document}

\title{The effect of geographic sampling on evaluation of extreme precipitation in high resolution climate models}

\author{Mark D. Risser and Michael F. Wehner}
\date{}
\maketitle

\begin{abstract}
{Traditional approaches for comparing global climate models and observational data products typically fail to account for the geographic location of the underlying weather station data. For modern high-resolution models, this is an oversight since there are likely grid cells where the physical output of a climate model is compared with a statistically interpolated quantity instead of actual measurements of the climate system. In this paper, we quantify the impact of geographic sampling on the relative performance of high resolution climate models’ representation of precipitation extremes in Boreal winter (DJF) over the contiguous United States (CONUS), comparing model output from five early submissions to the HighResMIP subproject of the CMIP6 experiment. We find that properly accounting for the geographic sampling of weather stations can significantly change the assessment of model performance. Across the models considered, failing to account for sampling impacts the different metrics (extreme bias, spatial pattern correlation, and spatial variability) in different ways (both increasing and decreasing). We argue that the geographic sampling of weather stations should be accounted for in order to yield a more straightforward and appropriate comparison between models and observational data sets, particularly for high resolution models. While we focus on the CONUS in this paper, our results have important implications for other global land regions where the sampling problem is more severe.}
\end{abstract}

%============================================
\section{Introduction} \label{section1}

Global climate models can contain significant uncertainties, particularly in their characterization of precipitation extremes. As a result, it is critical to use observationally-based data sets to evaluate a particular climate model to assess if the model is fit for purpose in exploring extremes and, if so, where and when the model is either acceptable or unacceptable for characterizing extreme precipitation. Traditionally, gridded daily products are used as a ``ground truth'' data set for evaluating a climate model because 
% \textbf
{(1) these data products are based on measurements of the real world (e.g., in situ measurements or satellite observations) and (2) they enable a like-for-like comparison between climate models and observations.
}

% Discuss the need to account for spatial sampling
% \textbf
{However,} for comparison with climate model output, the underlying physics of the climate model yields a process-based characterization of, e.g., extreme precipitation, at every grid cell while gridded products are based on spatially irregular measurements. %\textbf
{Consequently,} a comparison of the climate model versus a gridded product over an area with poor observational sampling (e.g., regions with large orographic variability) could be misleading, since the gridded product does not represent actual measurements of daily precipitation at these locations. 
In fact, this %\textbf
{issue} has already been examined in constructions of global mean temperature trends from station data \citep{madden1993bias, vose2005intercomparison}, although the effects of geographic sampling were minor partly because of the great care taken in the construction of trends \citep{jones2001adjusting}.

% Talk about high resolution models
This problem is likely to worsen when considering 
%\textbf
{very high resolution climate models, the evaluation of which is the primary motivation for this paper. For example, if one accounts for geographic sampling of the weather stations by only considering model grid cells with at least one representative high-quality station (the standard of comparison we use in this paper; see Section \ref{section32}), the resolution of even CMIP6-class models like CESM2 \citep{bacmeister2020co} and CanESM5 \citep{swart2019canadian} is coarse enough that a large majority grid cells over CONUS meet this criteria (in fact, all grid cells meet this criteria for CanESM5). However, when considering model output from five early submissions to the HighResMIP subproject of the CMIP6 experiment \citep{haarsma2016high}, the finer horizontal resolution means that at most 60\% of the grid cells meet this criteria over CONUS (see Table \ref{Ncells}). An extreme example is the HadGEM3-GC3.1-HM model \citep[][which has a $\sim$20km horizontal resolution]{Roberts_2019}, for which only 22\% of the model grid cells meet this criteria. In other words, for this particular selection criteria, accounting for geographic sampling with CanESM5 would have no impact on the model performance because all grid cells (about $\sim$300km across) have at least one high-quality station over the CONUS region. On the other hand, one might expect that the model performance could change drastically for the HadGEM3-GC3.1-HM model.
}

% Main points of the paper
% \textbf
{
In this paper, we make the case that geographic sampling of observational data should be taken into consideration when comparing climate models' representation of extremes to observations, particularly for high-resolution models. To this end, we develop a framework for systematically quantifying the effect of geographic sampling via a ``true'' standard of comparison based only on the model grid cells with a corresponding high-quality weather station. Our metrics for making this comparison are a measure of extreme bias and Taylor diagrams \citep{taylor2001summarizing}, which summarize the spatial pattern correlation and variability after removing any biases. For extreme precipitation in Boreal winter (DJF) over the contiguous United States (CONUS), we find that properly accounting for the geographic sampling of weather stations can significantly change the assessment of model performance. Across the models considered, failing to account for sampling impacts the different metrics in different ways (both increasing and decreasing). We argue that the geographic sampling of weather stations should be accounted for in order to yield a more straightforward and appropriate comparison between models and observational data sets. While we explore the contiguous United States (CONUS) in this paper, the geographic sampling issue is particularly important when considering global regions with very limited sampling (e.g., Central Asia, South America or Africa).
}

% Importance of pre-processing steps / order of operations
% Brief discussion of fractal scaling of precipitation / interpretation of model output
% \textbf
{
Before proceeding with our analysis, it is important to highlight several key points with respect to observational data products and the proper methodology for conducting model comparison for precipitation extremes. On one hand, in addition to the fact that daily gridded products are convenient for model comparison, it has been clearly documented that they are the best observational data source to use for yielding a like-for-like comparison with climate models for precipitation extremes \citep[e.g.,][]{chen2008verification,Gervais2014}. This is based on the fact that the correct interpretation for model grid cell precipitation is an areal average and not a point measurement \citep[see, e.g.,][]{chen2008verification}. Of course, weather stations yield point measurements, and so do daily gridded products that are based on an interpolation scheme or ``objective analysis'' \citep[e.g.,][]{livneh2015spatially}. However, when the native resolution of a gridded product is sufficiently finer than that of the climate model of interest, gridded point measurements can still be used for model evaluation of precipitation extremes so long as a specific workflow is followed. The proper steps are outlined in \cite{Gervais2014} and proceed as follows: (1) one must use a conservative remapping scheme for regridding the high-resolution gridded product \citep[e.g.,][which is flux-conserving]{jones2001adjusting}; and (2) one must calculate extreme statistics for comparison \textit{after} regridding the ``raw'' daily product.  If one instead (incorrectly) reverses this order of operations (i.e., calculate extreme statistics for the gridded product and then conduct the regridding), \cite{chen2008verification} demonstrate that any disagreement between models and observations could be solely due to grid size. 
}

% \textbf
{
On the other hand, as a purely observational product, a recent thread of research argues that gridded daily products are an inappropriate data source for characterizing pointwise measurements of extreme precipitation. The reasoning here is that daily precipitation is known to exhibit fractal scaling \citep[see, e.g.,][and numerous references therein]{Lovejoy2008,Maskey2016}, and therefore any spatial averaging during the gridding process will diminish variability and extreme values. There are an emerging number of analyses that specifically document this phenomena \citep{King2013,Gervais2014,Timmermans2018}; for example, \cite{Risser2019} show that daily gridded products underestimate long-period return values by 30\% or more, relative to in situ measurements. Gridded point-based extreme precipitation data products like \cite{donat2013updated} or \cite{Risser2019} preserve the extreme statistics of weather station measurements, but since they are not flux-conserving and furthermore do not account for the temporal occurrence of extreme events over space they can only provide local information about the climatology of precipitation extremes. Such products are very useful for characterizing the frequency of extreme precipitation for local impacts analyses, but are by construction an unsuitable data source for conducting model evaluation or comparison.
}

% Caveat: not doing model intercomparison / ranking
% \textbf
{
As a final note, we emphasize that in this paper we limit any comparisons to each individual model, focusing on the geographic sampling question, and specifically do not conduct intercomparisons with respect to ranking the models or providing general conclusions about their relative performance. Since high-resolution global climate models are the focus of this work, as previously mentioned we utilize output from several HighResMIP models with horizontal resolutions of $\sim 25$km to $\sim 50$km. However, the HighResMIP protocol \citep[outlined in][]{haarsma2016high} is unique in that it has been designed to systematically investigate the impact of increasing horizontal resolution in global climate models. To that end, the various modeling centers performed two simulations at two spatial resolutions for each model. The HighResMIP protocol recommends that only the lower resolution version of the model is tuned and that the same set of parameters is used, as far as possible, in the simulation at high spatial resolution. Therefore, the high-resolution simulations (which are the simulations used in this study) have not been designed to be the best possible simulations of each individual model, even if they have generally been performed using the latest version of each model. This is particularly important for precipitation extremes that are strongly influenced by moist physics parametrizations. 
}

% Paper outline
The paper proceeds as follows: in Section~\ref{section2} we describe the various data sources used (gridded daily products and climate model output) in our analysis, and in Section \ref{StatMethods} we describe the statistical methods %\textbf
{and framework for accounting for geographic sampling}. In Section~\ref{sec:results}, we illustrate our methodology using a case study comparing a well-sampled (spatially) region versus a poorly-sampled region before presenting the results of our analysis for all of CONUS, maintaining a focus on Boreal winter (DJF) precipitation. Section~\ref{sec:discussion} concludes the paper.

%============================================
\section{Data sources} \label{section2}

\subsection{Observational reference data}

% \textbf
{As described in Section \ref{section1}, model simulated precipitation is best interpreted as an areal average over the model grid cell \citep{chen2008verification, Gervais2014}, as opposed to the point measurement interpretation appropriate to in situ weather station measurements. Following \cite{Gervais2014}, the correct way to compare model data with station data for precipitation extremes is as follows:
\begin{enumerate}
\item Use an objective analysis to translate daily station (point) measurements to a grid with a much higher resolution than the model resolution,
\item Still working with the daily data, use a conservative remapping procedure \citep[e.g.,][]{jones1999first} to translate the high-resolution grid values to the climate model grid, and finally
\item Calculate extreme statistics of interest on the common grid and compare.
\end{enumerate}
For step 2 it is critical to use a conservative remapping procedure, since this is consistent with the ``areal average'' interpretation of model grid cell precipitation \citep{Gervais2014}. Alternatively, any interpolation scheme is only appropriate for the ``point measurement'' interpretation of precipitation. Given that daily precipitation is known to exhibit fractal scaling \citep[see, e.g.,][and numerous references therein]{Lovejoy2008,Maskey2016} the objective analysis in step 1 will diminish variability and extreme values \citep{King2013,Gervais2014,Timmermans2018,Risser2019}. Nonetheless, applying conservative remapping to gridded daily precipitation products is a flux-conserving operation, and the resulting data are hence the right source to use for model comparison.}

% \textbf
{
One pre-existing gridded product that meets the criteria of step 1 above is the \cite{livneh2015spatially,livneh2015dataset} daily gridded product (henceforth L15), which has been gridded to a $1/16^\circ$ or $\sim6$km horizontal resolution and spans the period 1950-2013. While the data product covers North America (south of $53^\circ$N), we limit our consideration to those grid cells within the boundaries of the contiguous United States (CONUS). The L15 data product takes in situ measurements of daily total precipitation \citep[over CONUS, the input data are from the Global Historical Climatology Network;][]{Menne2012} and creates a daily gridded product in two steps. First, for each day, the station measurements are interpolated to a $1/16^\circ$ high-resolution grid using the SYMAP algorithm \citep{shepard1968two,shepard1984computer}, which is an inverse-distance weighting approach that assigns weights to each grid cell only based on nearby points (for computational feasibility) and accounts for directionality. Second, the interpolated data are multiplied by a monthly scaling factor that is determined by the ratio of its mean monthly baseline climatology (1981-2010) and the mean monthly climatology from the same period of the topographically-aware PRISM  data product \citep{daly1994statistical, daly2008physiographically}. The scaling factor is designed to adjust the ``topographically unaware'' interpolated data to scale meaningfully with orography. While the combination of interpolating point measurements and applying a monthly rescaling likely has an impact on observed extreme events, such considerations are beyond the scope of this paper.
}

% \textbf
{
Following the remaining steps outlined by \cite{Gervais2014}, for each of the climate model grids considered (see Table \ref{modelDetails}) we first regrid the daily L15 data to the model grid using a conservative remapping procedure (we utilize functionality from the \pkg{rainfarmr} package for \proglang{R}; \citealp{R_rainfarmr}). Then, for each grid cell, we then extract the largest running 5-day precipitation total (denoted Rx5Day) in DJF, denoted
\[
\big\{ Y_{t}^{L,m}(\bfg): t = 1951, \dots, 2013; \bfg \in \mathcal{G}_m \big\},
\]
where $\mathcal{G}_m$ is the model grid for model $m = 1, \dots, 5$.
}

% \textbf
{
The L15 data product is, of course, only one of a very large number of gridded daily precipitation products that could be considered in this study. However, we choose to use L15 for several reasons: first, it is one of the more widely used gridded daily products; second, it covers a relatively long time record (64 years); and finally, its native resolution is sufficiently higher than the climate models considered in this paper such that it can be conservatively remapped to the model grids following the procedure outlined in \cite{Gervais2014}. The Climate Prediction Center (CPC) $0.25^\circ \times 0.25^\circ$ Daily US Unified Gauge-Based Analysis of Precipitation is another commonly used data product with similar time coverage; however, its $0.25^\circ$ resolution is approximately the same as the models considered in this study. Since the CPC product represents a point measurement, its resolution is too coarse to be appropriately translated to an areal average for use in HighResMIP model evaluation according to the \cite{Gervais2014} methodology.
}

\subsection{Climate models} \label{climMod}

\begin{table}[t]
    \centering
    \begin{tabular}{|c|c|c|c|c|}
    \hline
    \textbf{Model} & \textbf{Label} & \textbf{Resolution} & \textbf{Ensemble members} & \textbf{CONUS grid cells} \\ \hline \hline
        CNRM-CM6-1-HR$^\dagger$ & CNRM   & $\approx$0.5$^\circ$  & 1 & 3256 \\ \hline
        ECMWF-IFS-HR & ECMWF      & $\approx$0.5$^\circ$  & 4 & 3253 \\ \hline
        HadGEM3-GC31-HM & HadGEM  & $\approx$0.25$^\circ$ & 3 & 9900 \\ \hline
        IPSL-CM6A-ATM-HR & IPSL   & $\approx$0.5$^\circ$  & 1 & 2316 \\ \hline
        MPI-ESM1-2-XR & MPI   & $\approx$0.5$^\circ$  & 1 & 3748 \\ \hline
    \end{tabular}
    \caption{HighResMIP climate model output used in the analyses for this paper. $^\dagger$Note: the CNRM runs are only from 1981-2014.}
    \label{modelDetails}
\end{table}

Given that this study is motivated by the evaluation of high-resolution climate models, we utilize early submissions to the highresSST-present experiment of the HighResMIP subproject of the CMIP6 experiment \citep{haarsma2016high}, all of which are AMIP-style runs with fixed sea surface temperatures from 1950-2014. 
% \textbf
{As mentioned in Section \ref{section1}, the HighResMIP protocol was designed to systematically explore the impact of increasing horizontal resolution in global climate models. Each of the models used here was run at two spatial resolutions, including a high spatial resolution. However, the protocol specifies that only the low-resolution version of the model is tuned and that the same set of parameters is used for the high-resolution simulations. We utilize only the high-resolution simulations, even though several of the models have scale-aware parameterizations (i.e., they have to be modified with increasing resolution) and therefore there might be large differences between the models in terms of the tuning used in the simulations evaluated here. For this reason, we reiterate that in this paper we maintain comparisons to within-model statements and do not attempt to conduct intercomparison.
}

The following models are used for our analysis:
\begin{enumerate}
    \item CNRM-CM6-1-HR: This model is developed jointly by the CNRM-GAME (Centre National de Recherches M\'et\'eorologiques--Groupe d’\'etudes de l’Atmosph\`ere M\'et\'eorologique) in Toulouse, France and CERFACS (Centre Europ\'een de Recherche et de Formation Avanc\'ee); see  \cite{Voldoire2013}. Data is interpolated to $0.5^\circ$ regular latitude longitude grid from the native T359l reduced Gaussian grid. The model has 91 vertical levels with the top level at 78.4km.
    \item ECMWF-IFS-HR: The Integrated Forecasting System (IFS) model of the European Centre for Medium-range Weather Forecasting as configured for multi-decadal ensemble climate change experiments \cite{roberts2018climate}. Data is interpolated onto a $0.5\times 0.5$ regular latitude-longitude grid from the native Tco399 cubic octahedral reduced Gaussian grid (nominally $\sim$25km). The model has 91 vertical levels with the top model level at 1 hPa. 
    \item HadGEM3-GC31-HM: The UK MetOffice Hadley Centre (Exeter, United Kingdom) unified climate model, HadGEM3-GC3.1-HM on a regular 768$\times$1024 latitude-longitude grid \citep[nominally 25km;][]{Roberts_2019} The model has 85 vertical levels with the top model level at 85 km. 
    \item IPSL-CM6A-ATM-HR: This model is developed by the Institut Pierre Simon Laplace in  Paris, France \citep{IPSL_citation}. Data is interpolated from the native N256 geodesic grid to a 512 $\times$ 360 longitude-latitude grid  (nominally $\sim$50km). The model has 79 vertical levels with the  top model level at 40000 m.
    \item MPI-ESM1-2-XR: The MPI-ESM1-2-XR from the Max Planck Institute for Meteorology in Hamburg, Germany \citep{gutjahr2019max}. Data is interpolated to a regular 768$\times$384 longitude/latitude (nominally $\sim$50km) from the native T255 spectral grid. The model has 95 vertical levels with the top model level at 0.1hPa.
\end{enumerate}
For each of these models (additional details provided in Table \ref{modelDetails}) and each ensemble member, we calculate the corresponding DJF Rx5Day values for each ensemble member in each grid box over CONUS. Furthermore, note that we mask out grid cells that are not fully over land. These values are denoted 
\[
\big\{ Y_{t,e}^{m}(\bfg): t = 1951, \dots, 2014; \bfg \in \mathcal{G}_m; e = 1, \dots, N_{\text{ens},m}\big\}, 
\]
for model $m = 1, \dots, 5$, where $\mathcal{G}_m$ is the model grid for model $m$ and $N_{\text{ens},m}$ is the number of ensemble members for model $m$ (see Table \ref{modelDetails}); however, note that the CNRM runs only cover 1981-2014.

%============================================
\section{Methods} \label{StatMethods}

\subsection{Extreme value analysis} \label{spatialGEV}

% \textbf
{
The core element of our statistical analysis is to estimate the climatological features of extreme precipitation for the regridded L15 and each climate model using the Generalized Extreme Value (GEV) family of distributions, which is a statistical modeling framework for the maxima of a process over a pre-specified time interval or ``block,'' e.g., the three-month DJF season used here.}
\cite{Coles2001} (Theorem~3.1.1, page~48) shows that when the number of measurements per block is large, the cumulative distribution function (CDF) of %\textbf
{the seasonal Rx5Day} $Y_t(\bfg)$ %\textbf
{can be approximated by a} member of the GEV family
\begin{equation} \label{gev_fam}
G_{\bfg, t}(y) \equiv \mathbb{P}(Y_t(\bfg) \leq y) = \exp\left\{-\left[ 1 + \xi_t(\bfg)\left(\frac{y - \mu_t(\bfg)}{\sigma_t(\bfg)}\right) \right]^{-1/\xi_t(\bfg)} \right\}, 
\end{equation}
defined for $\{ y: 1 + \xi_t(\bfg)(y - \mu_t(\bfg))/\sigma_t(\bfg) > 0 \}$. The GEV family of distributions~(\ref{gev_fam}) is characterized by three space-time parameters: the location parameter $\mu_t(\bfg) \in \mathcal{R}${, }which describes the center of the distribution{;} the scale parameter $\sigma_t(\bfg)>0$, which describes the spread of the distribution{;} and the shape parameter $\xi_t(\bfg) \in \mathcal{R}$. The shape parameter $\xi_t(\bfg)$ is the most important {for} determining the qualitative behavior of the distribution of daily rainfall at a given location{. I}f $\xi_t(\bfg)<0$, the distribution has a finite upper bound; if $\xi_t(\bfg) >0$, the distribution has no upper limit; {and} if $\xi_t(\bfg) = 0$, the distribution is again unbounded and the CDF~(\ref{gev_fam}) is interpreted as the limit $\xi_t(\bfg) \rightarrow 0$ \citep{Coles2001}. %\textbf
{While the GEV distribution is only technically appropriate for seasonal maxima as the block size approaches infinity,}
%% Talk about the appropriateness of the GEV for finite block sizes.
\cite{risser2019detected} verify that the GEV approximation is appropriate %\textbf
{here even though the number of ``independent'' measurements of Rx5Day in a season is relatively small}, particularly when limiting oneself to return periods within the range of the data (as we consider here, with the 20-year return values).

While our goal is to simply estimate the climatology of extreme precipitation (and specifically not to estimate or detect trends), the nonstationarity of extreme precipitation over the last 50 to 100 years \citep[see, e.g.,][]{Kunkel2003,min2011human,zhang2013attributing,fischer2015anthropogenic,Easterling2017,risser2019detected} requires that we characterize a time-varying extreme value distribution. Here, we use the simple trend model
\begin{equation} \label{coef_model}
\mu_t(\bfg) = \mu_0(\bfg) + \mu_1(\bfg) X_t, \hskip2ex \sigma_t(\bfg) \equiv \sigma(\bfg), \hskip2ex \xi_t(\bfg) \equiv \xi(\bfg),
\end{equation}
%\textbf
{where $X_t = [\text{GMT}]_t$ is the smoothed (5-year running average) annual global mean temperature anomaly in year $t$, obtained from GISTEMPv4 \citep{lenssen2019improvements,GISSTEMPv4}. The global mean temperature is a useful covariate for describing changes in the distribution of extreme precipitation, although other process based covariates could work equally well or better \citep{Risser2017}.} While this is an admittedly simple temporal model, we argue that it is sufficient for characterizing the climatology of seasonal Rx5Day. Furthermore, \cite{risser2019detected} use a similar trend model as (\ref{coef_model}), which they show to be as good (in a statistical sense) as more sophisticated trend models (where, e.g., the scale and/or shape vary over time). 
% \textbf
{While it has been argued that much more data is required to fit non-stationary models like (\ref{coef_model}) reliably \citep{li2019much}, the inclusion of a single additional statistical parameter is used to address the fact that seasonal Rx5Day over CONUS is not identically distributed over 1950-2013. Furthermore, all comparisons in this paper are based on a time-averaged return value and we do not attempt to directly interpret any temporal changes in the distribution of Rx5Day.}
We henceforth refer to $\mu_0(\bfg)$, $\mu_1(\bfg)$, $\sigma(\bfg)$, and $\xi(\bfg)$ as the \textit{climatological coefficients} for grid cell $\bfg$, as these values describe the climatological distribution of extreme precipitation in each year. 

% \textbf
{For each model grid and data type (climate model output or regridded L15), we utilize the \pkg{climextRemes} package for \proglang{R} \citep{R_climextRemes} to obtain maximum likelihood estimates (MLEs) of the climatological coefficients, denoted
\begin{equation} \label{smoothEst}
\{ \widehat{\mu}_0(\bfg), \widehat{\mu}_1(\bfg), \widehat{\sigma}(\bfg), \widehat{\xi}(\bfg) \},
\end{equation}
independently for each grid cell. Within \pkg{climextRemes}, we also utilize the block bootstrap \citep[see, e.g.,][]{Risser2019} to quantify uncertainty in the climatological coefficients in all data sets considered. For those models with more than one ensemble member (see Table \ref{modelDetails}), we treat the ensemble members as replicates and obtain a single MLE of the climatological coefficients.}
The MLEs from (\ref{smoothEst}) %\textbf
{and the bootstrap estimates} can be used to calculate corresponding estimates of the DJF climatological $20$-year return value, denoted $\widehat{\phi}(\bfg)$, which is defined as the DJF maximum five-daily precipitation total that is expected to be exceeded on average once every $20$ years in grid cell $\bfg$ under a fixed GMT anomaly. In other words, $\widehat{\phi}(\bfg)$ is an estimate of the $1-1/20$ quantile of the distribution of DJF maximum five-daily precipitation at grid cell $\bfg$, i.e.,
$P\big(Y_{t}(\bfg) > \widehat{\phi}(\bfg)\big) = {1}/{20}$, which can be written in closed form in terms of the climatological coefficients:
\begin{equation} \label{returnVal}
\widehat{\phi}(\bfg) = \left\{ \begin{array}{ll}
[\widehat{\mu}_0(\bfg) + \widehat{\mu}_1(\bfg) \overline{X}] - \frac{\widehat{\sigma}(\bfg)}{\widehat{\xi}(\bfg)}\big[1 - \{-\log(1-1/r)\}^{-\widehat{\xi}(\bfg)}\big],  & \widehat{\xi}(\bfg) \neq 0 \\[1ex]
[\widehat{\mu}_0(\bfg) + \widehat{\mu}_1(\bfg)\overline{X}] - \widehat{\sigma}(\bfg) \log\{-\log(1-1/r)\},  & \widehat{\xi}(\bfg) = 0
\end{array} \right. 
\end{equation}
\citep{Coles2001}; %\textbf
{here $\overline{X}$ is the average GMT anomaly over 1950 to 2013. Averaging over the GMT values yields a time-averaged or climatological estimate of the return values.} 

% \textbf
{At the end of this procedure, we have MLEs of the climatological 20-year return value $\widehat{\phi}(\bfg)$ for each of the five climate models and each of the five regridded L15 data sets, as well as bootstrap estimates of these return values $\{ \widehat{\phi}_b(\bfg): b = 1, \dots, 250\}$ for each model grid and data type.}

\subsection{Comparing the climatology of extreme precipitation} \label{section32}

We illustrate the effect of geographic sampling on the evaluation of simulated 20-year return values of winter (DJF) maximum five-day precipitation from selected high resolution climate models with  %\textbf
{Taylor diagrams \citep{taylor2001summarizing} to illustrate pattern errors and return value bias to quantify magnitude errors}. 
Taylor diagrams \citep{taylor2001summarizing} plot the centered pattern correlation between observations and simulations as the angular dimension and the ratio of the observed to simulated spatial standard deviation as the radial dimension. These diagrams provide information about the spatial pattern of model errors with the biases removed.  The bias in 20-year return values (hereafter referred to as ``extreme bias''), %\textbf
{defined as the absolute difference between the return values from model and observations,} provides a simple measure of whether the models are too dry or too wet, while the Taylor diagram provides three useful metrics displayed in a single plot: (1) the spatial pattern correlation between the model and observations, (2) a comparison of the spatial standard deviation over the region, and (3) a skill score to assess a level of agreement between the two spatial fields. Taylor's modified skill score $\mathcal{S}$, comparing two spatial fields (e.g., a model versus observations), is defined as
\[
\mathcal{S} = \exp\left\{ -\frac{s_1^2 + s_2^2 -2s_1s_2r }{2s_1s_2}  \right\},
\]
where $s_j$ is the standard deviation of spatial field $j=1,2$ and $r$ is the %\textbf
{spatial pattern} correlation between the two fields \citep[also used in][]{wehner2013very}. The skill score essentially involves the ratio of the mean squared error between the two fields (after removing the average from each field) and the standard deviations of each field; the score $\mathcal{S}$ ranges between 0 (indicating low skill) and 1 (indicating perfect skill).  Further details on the Taylor diagrams are provided in Section \ref{sec:results}; however, an important detail is that to calculate a Taylor diagram we must have ``paired'' observations and model data, i.e., both data sources must be defined on the same spatial support or grid, 
% \textbf
{which fits nicely into the framework considered in this paper}. 

% \textbf
{
In order to illustrate the effect of geographic sampling, we can compare the extreme bias and Taylor diagrams results for two approaches:}
\begin{enumerate}
\item[A1] {``True'' model performance}

{
First, in what we regard as the ``true'' performance for each model, we calculate the extreme bias and Taylor diagram metrics using the subset of model grid cells (for the model and regridded L15) that have corresponding weather station measurements. L15 is based on measurements from the GHCN-D \citep{Menne2012} records, but its stability constraint \citep[selecting stations with a minimum of 20 years of data over CONUS;][]{livneh2015spatially} means that the specific station measurements that go into the daily gridded product change over time. To navigate this complication, we define a set of ``high-quality'' stations from the GHCN to be those stations that have at least 90\% non-missing daily measurements over 1950-2013. This results in $n = 2474$ stations with a relatively good spatial coverage of CONUS (see Figure \ref{hqGHCNstations}), although the coverage is much better in the eastern United States relative to, e.g., the Mountain West. The 90\% threshold is somewhat arbitrary, but this cutoff ensures that these stations enter into the L15 gridding procedure for a large majority of days over 1950-2013. Next, for each model grid, we identify the grid cells that have at least one high-quality GHCN station (see Figure \ref{model_masks} in the Appendix). The extreme bias and Taylor diagram are then calculated for only those grid cells with at least one high-quality station. Since this comparison isolates the regridded L15 cells that actually involve a real weather station measurement, we regard this as the true extreme bias and Taylor diagram metrics for each model.}

\item[A2] {Ignore geographic sampling}

{To assess how the performance of each model changes when the geographic sampling of weather stations is not accounted for, we calculate the extreme bias and Taylor diagram metrics using all grid cells.}

\end{enumerate}
% \textbf
{Approach A1 provides us with a standard of comparison, since it properly accounts for the geographic sampling of the underlying weather stations, and we argue that this is the most appropriate way to conduct model comparison. Approach A2, on the other hand, is what would be done without consideration for the geographic sampling issue. Comparing the extreme bias and Taylor diagrams for A1 versus A2 allows us to explicitly quantify the effect of geographic sampling on assessed  model performance.
}

%============================================
\section{Results} \label{sec:results}

% \textbf
{
For reference, we have provided a supplemental figure for each model that shows the estimated 20-year return values for the climate model output and regridded L15, the corresponding bootstrap standard errors, the absolute difference in return values, and the ratio of standard errors for the climate model vs. regridded L15; see Figures \ref{CNRM}, \ref{ECMWF}, \ref{HadGEM}, \ref{IPSL}, and \ref{MPI}.
}

\subsection{Case study: Kansas versus Utah} \label{section41}

\begin{figure}[!t]
\begin{center}
\includegraphics[trim={0 0 0 0mm}, clip, width = 0.95\textwidth]{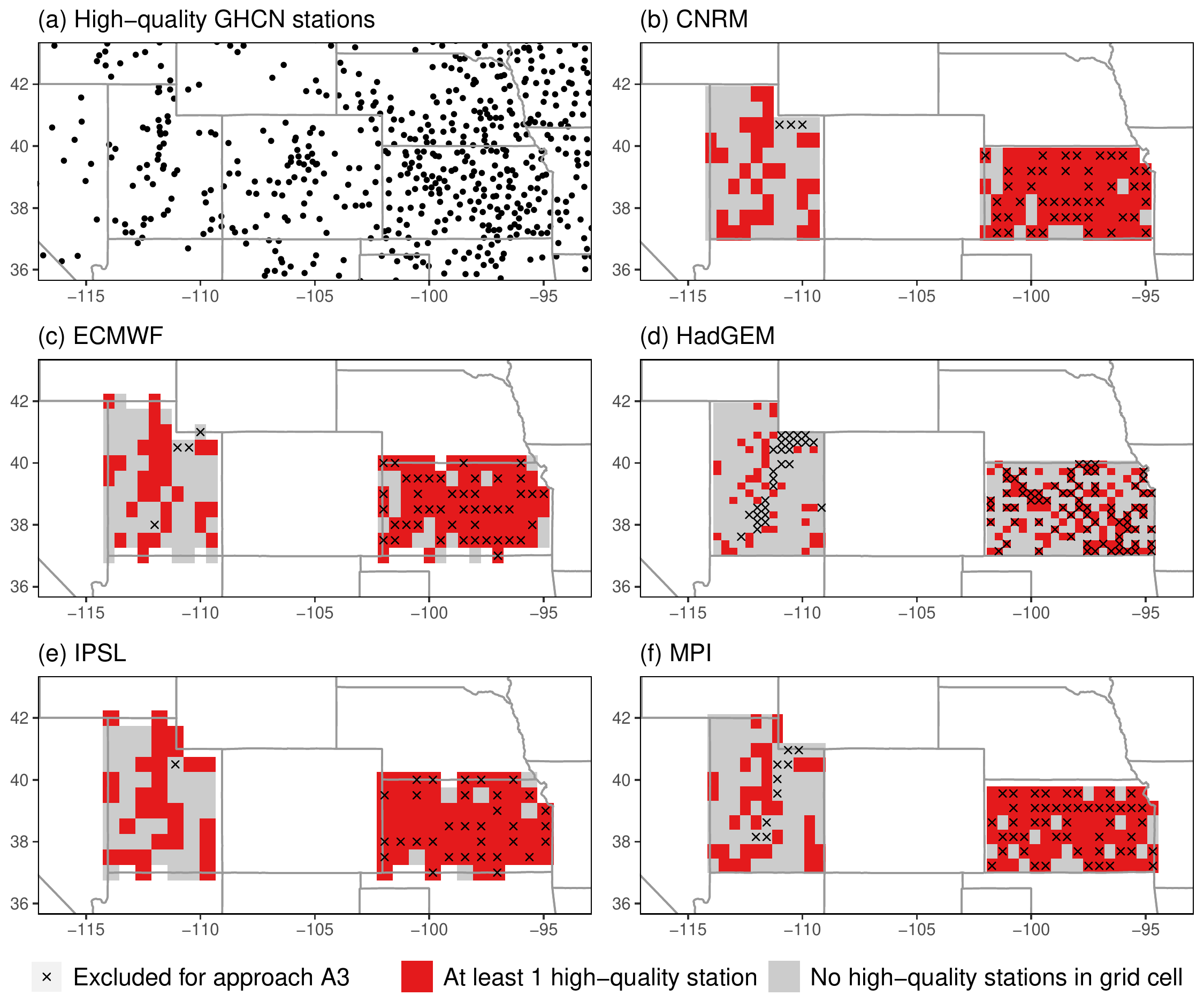}
\caption{Model grid cells with at least one high-quality GHCN station for Kansas and Utah, with an $\times$ denoting model grid cells that are excluded for A3$_\text{KS}$ and A3$_\text{UT}$.} 
\label{modelMask_KS_UT} 
\end{center}
\end{figure}

% \textbf
{
To illustrate our methodology, we first explore return value estimates for two small spatial subregions that exhibit very different sampling by the GHCN network, namely Kansas and Utah. These two states present an illustrative case study for our method because Utah is poorly sampled (52 stations over 219,890 km$^2$), while Kansas is well sampled (140 stations over 213,100 km$^2$). Figure \ref{modelMask_KS_UT} shows the model grid cells with and without at least one high-quality station. In Kansas, there are anywhere from 40\% (HadGEM) to 90\% (IPSL) of the model grid cells with a high-quality station; in Utah, between 16\% (HadGEM) and 45\% (IPSL) of the model grid cells have a high-quality station (see Table \ref{Ncells} in the Appendix). The L15 climatology of 20-year return values is quite similar in these two states: the median return value (inter-quartile range) in Kansas is 53.8mm ($[39.4, 61.0]$), while in Utah the median is 49.6mm ($[34.3, 71.1]$).
}

% \textbf
{However, Utah also differs markedly from Kansas with respect to topography: Kansas is relatively flat while Utah exhibits complex orographic variability. While the geographic sampling of stations is likely decoupled from rainfall behavior in Kansas, in Utah stations are primarily located at lower elevations where extreme orographic precipitation does not occur. Indeed, the median elevation of the high-quality stations in Utah is around 1600m (the highest station is at 2412m), while the highest peaks in Utah exceed 4000m. When considering the model grids, the geographic sampling of the weather stations excludes the highest elevations for all models except IPSL, which has the coarsest resolution of those considered in this paper. For example, looking at the relationship between the average elevation of each grid cell (averaged from the GTOPO30 1km digital elevation data set) and the 20-year return values for both the model output and L15 (see Figure \ref{elev_vs_rv} in the Appendix), it is clear that grid cells with the highest elevations in Utah do not have a corresponding high-quality station. On the other hand, this phenomena is nearly irrelevant for Kansas (as expected), since the unsampled model grid cells mostly fall within the elevation range of the sampled cells. 
}

% \textbf
{The implication here is that any relative differences between approaches A1 and A2 for Kansas versus Utah could be due to either the different geographic sampling of the two states or orographic considerations, or both. To explicitly separate these two possibilities, we introduce a third approach for the case study, denoted A3. Given the differences in Kansas and Utah (with respect to sampling density and orography), we define this approach differently for each state:
}
\begin{itemize}
\item[A3$_\text{KS}$] {Consider a subset of grid cells with a high-quality station}

% \textbf
{The difference between A1 and A2 in Kansas is going to be minimal for many of the models simply because the state is extremely well sampled (CNRM, ECMWF, IPSL, and MPI all have greater than 80\% of their model grid cells with a representative high-quality station). To assess the influence of the high station density in Kansas, we can alternatively compare the performance of each model when we randomly subsample the model grid cells such that the proportion of cells with a high-quality station matches that of Utah. For example, for CNRM, only $33/92 \approx 0.36$ of the model grid cells have at least one high-quality station in Utah (see Table \ref{Ncells} in the Appendix); we can randomly select 32 of the 75 CNRM grid cells in Kansas that have a high-quality station so that the proportion of Kansas sampled matches that of Utah ($32/89 \approx 0.36$). In other words, this approach allows us to answer the question \textit{``What would the effect of geographic sampling be for Kansas if its sampling density matched Utah?"} The grid cells excluded for A3$_\text{KS}$ are shown in Figure \ref{modelMask_KS_UT}.
}

\item[A3$_\text{UT}$] {Ignore geographic sampling but threshold high elevations}

% \textbf
{For Utah, this approach ignores geographic sampling (like A2) but only considers those model grid cells whose elevation does not exceed the highest grid cell with a high-quality station. For example, for HadGEM, the A3$_\text{UT}$ approach excludes any model grid cell with an average elevation that exceeds 2380m (which is the highest grid cell with a high-quality station for the HadGEM grid; see Figure \ref{elev_vs_rv} in the Appendix). In other words, this approach allows us to answer the question \textit{``What would the effect of geographic sampling be for Utah if we remove systematic differences in orography?"} The grid cells excluded for A3$_\text{UT}$ are also shown in Figure \ref{modelMask_KS_UT}.
}
\end{itemize}

\begin{figure}[!t]
\begin{center}
\includegraphics[trim={0 0 0 0mm}, clip, width =\textwidth]{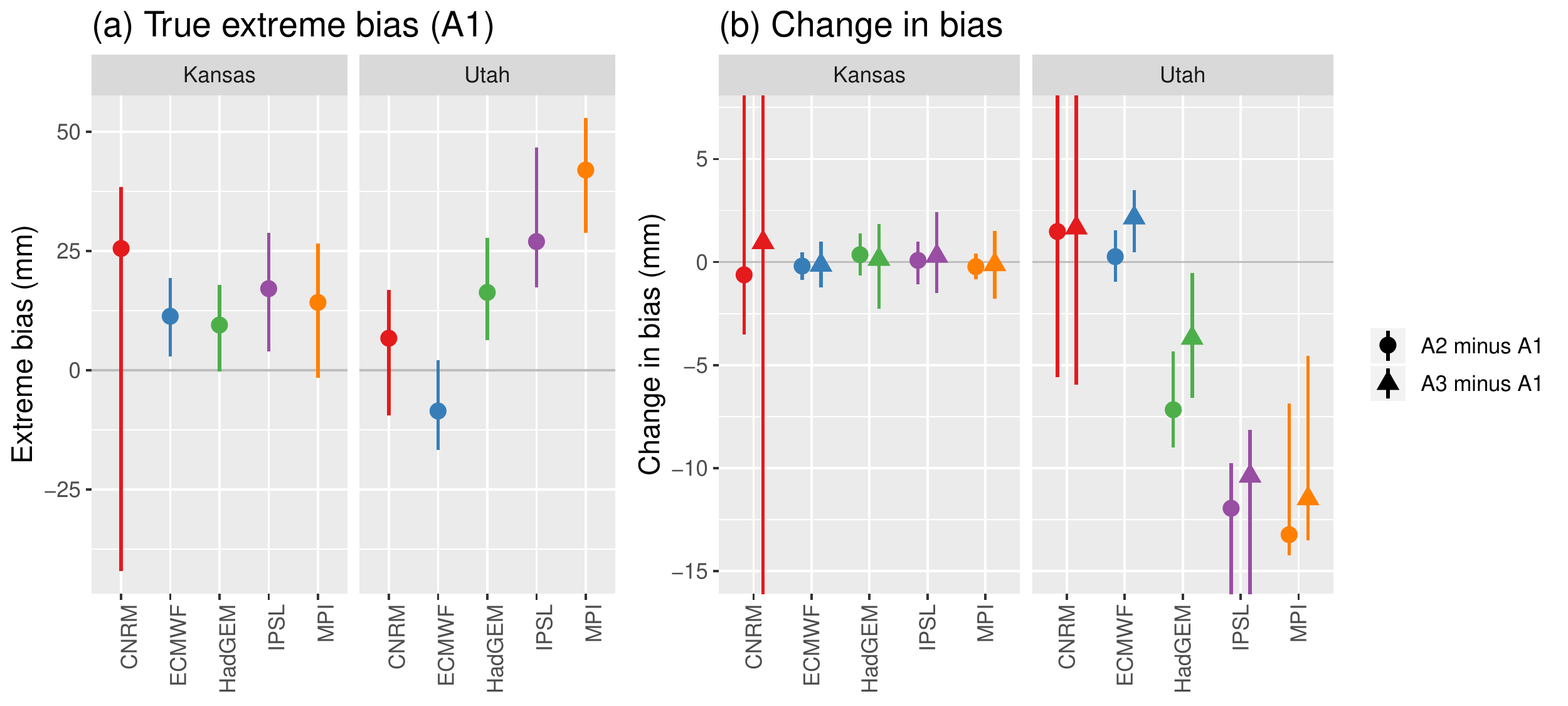}
\caption{True extreme bias (difference in 20-year return values) in Kansas and Utah for each model (approach A1; model minus regridded L15), with the change in extreme bias (in mm) for approach A2 (ignoring geographic sampling) versus A1 and approach A3 (assessing the effect of sampling density for Kansas and orography for Utah) versus A1. All estimates show the 95\% basic bootstrap confidence intervals.}
\label{bias_utahkansas} 
\end{center}
\end{figure}

% \textbf
{The true extreme bias averaged over each state with a 95\% basic bootstrap confidence interval (CI) is shown in Figure \ref{bias_utahkansas}(a). In Kansas, the models are uniformly too wet, with positive biases of approximately 5-20mm in all models and only the CNRM CI significantly overlapping zero. The uncertainty for CNRM is significantly larger than the other models in Kansas, which may be related to the fact that these runs only cover 1981-2014 while the other models cover 1950-2014. In Utah, on the other hand, the models are generally too wet again, although the biases for IPSL and MPI are particularly large. The ECMWF model's dry bias is a notable exception in Utah. The model biases for Kansas are all within each other's uncertainties, while there are meaningful differences in the true bias across the models for Utah. 
}

% \textbf
{
Turning to the change in bias in Figure \ref{bias_utahkansas}(b), there is a minimal effect of geographic sampling in Kansas: the CIs for the difference in extreme bias for A2 versus A1 include zero for all models. Interestingly, this remains true for approach A3$_\text{KS}$, where the CIs for A3$_\text{KS}$ versus A1 still include zero and completely overlap the CIs for A2 versus A1 (although the uncertainty in the A3$_\text{KS}$ versus A1 difference is extremely large for CNRM). 
The story is much different for Utah, where the CIs for A2 versus A1 do not include zero for HadGEM, IPSL, and MPI. The implication is that for these models, failing to account for geographic sampling makes each model appear to be drier than it actually is (even though their true biases are positive, i.e., too wet). For CNRM and ECMWF, on the other hand, the best estimate of the difference between A2 and A1 is positive, meaning that ignoring geographic sampling makes the models appear to be too wet (however, their confidence intervals include zero). Considering the A3$_\text{UT}$ versus A1 changes in extreme bias, there are interesting differences for all models except CNRM. In ECMWF the extreme bias gets larger, which is unusual (although the A2 vs. A1 and A3$_\text{UT}$ vs. A1 CIs overlap), while the extreme biases decrease in absolute value for HadGEM, IPSL, and MPI. For IPSL and MPI these decreases are not meaningful (the CIs nearly coincide with one another), and while the HadGEM CIs also overlap the changes are larger. This might have something to do with model resolution, since the HadGEM model has the highest resolution of those considered in this study.
}

% \textbf
{These two states provide important insights into the effect of geographic sampling with respect to the extreme bias. In Kansas, the changes in extreme bias are nonsignificant even when we artificially reduce the amount of information from the underlying weather stations. The implication is that for relatively homogeneous domains like Kansas the geographic sampling and its density is less important: in other words, to accurately evaluate models, fewer stations are required in topographically flat regions. In Utah, the geographic sampling is much more important since there are significantly non-zero changes to the extreme bias when it is ignored. Accounting for the influence of orography decreases this effect (i.e., the change in bias is generally less for A3$_\text{UT}$ relative to A2), but the interesting point is that when sampling matters (i.e., when the A2 vs. A1 confidence interval does not include zero) it \textit{still} matters regardless of whether accounting for elevation or not. A possible exception is ECMWF, for which the A2 vs. A1 CI includes zero while the A3$_\text{UT}$ vs. A1 CI does not, but there are clearly systematic differences between ECMWF and the other models considered in Utah based on its true bias.}

\begin{figure}[!t]
\begin{center}
\includegraphics[trim={0 0 0 0mm}, clip, width =\textwidth]{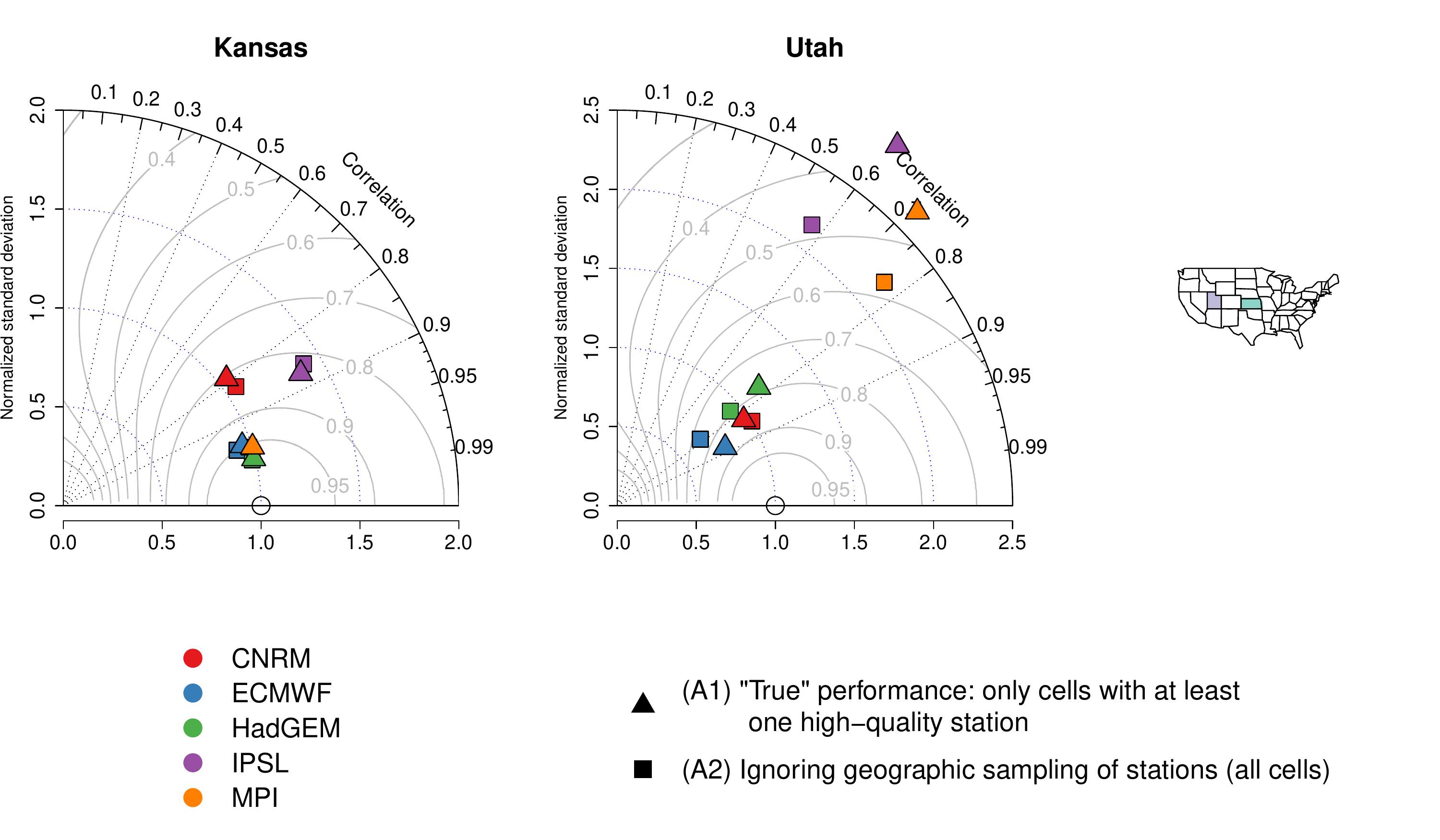}
\caption{Taylor diagrams comparing 20-year return values for Kansas and Utah for the climate models versus regridded L15. The gray curvilinear lines represent the skill score for each model relative to the regridded L15.}
\label{taylorDiag_utahkansas} 
\end{center}
\end{figure}

% \textbf
{The Taylor diagrams comparing 20-year return values for Kansas and Utah are shown in Figure \ref{taylorDiag_utahkansas}. In light of the results for the extreme bias and in order to not complicate the plot, we have chosen to leave the A3 comparison off of the Taylor diagrams (furthermore, we do not show the uncertainty in the Taylor diagrams, again for simplicity). Clearly, as with the extreme bias, the performance of the various climate models in replicating the spatial pattern and variability of extreme precipitation is very different for these two states. In Kansas, all models except IPSL almost perfectly reproduce the spatial variability of the return values, with or without accounting for geographic sampling. ECMWF, HadGEM, and MPI furthermore nearly reproduce the spatial patterns of extreme precipitation (again with or without accounting for geographic sampling), with spatial pattern correlations in excess of 0.95. Across all models the performance is nearly identical regardless of whether the geographic sampling of stations is taken into consideration. In Utah, on the other hand, for all models except CNRM there appears to be a noticeable difference when ignoring geographic sampling. This is particularly true for IPSL and MPI: the difference in spatial variability is significantly larger when accounting for geographic sampling. Interestingly, the spatial pattern correlation is roughly the same for A1 and A2 in IPSL and MPI, but the skill scores are lower when accounting for geographic sampling. The implication is that, at least with respect to the skill scores, failing to account for geographic sampling would lead one to conclude that IPSL and MPI perform better than they actually do. The differences in spatial variability, pattern correlation, and skill score are much smaller for the other three models in Utah, although it is the case that accounting for geographic sampling slightly improves the skill score for ECMWF.
}

% \textbf
{In summary, the main points of this case study are as follows: for well-sampled regions (like Kansas), the extreme bias, spatial pattern correlation, and spatial variability are approximately the same regardless of whether geographic sampling is explicitly accounted for, while for poorly sampled regions (like Utah) the geographic sampling can have an unpredictable impact on the extreme bias and Taylor diagram metrics. These results hold true even when reducing the sampling density in a well-sampled region (Kansas) and when accounting for the effect of extreme orography (in Utah), in the sense that when geographic sampling matters for a topographically heterogeneous region it still matters even after ignoring unsampled elevations. And, critically, it is important to note that while failing to account for geographic sampling changes the models' performance, it does not do so systematically: in Utah, the extreme bias both increases (ECWMF) and decreases (HadGEM, IPSL, and MPI); the skill scores both increase (ECMWF) and decrease (IPSL and MPI).
}

\subsection{Comparisons for CONUS and large climate subregions} \label{section42}

To explore these considerations more broadly, we now expand the scope of our model evaluation to systematically consider all of CONUS and seven spatial subregions. These subregions (shown in Figure \ref{taylorDiags} and also with labels in Figure \ref{FigA4} in the Appendix) are loosely based upon the climate regions used in the National Climate Assessment \citep{wuebbles2017climate}, with a small adjustment in the western United States to make the regions somewhat homogeneous with respect to the geographic sampling of the GHCN stations. Table \ref{Ncells} in the Appendix summarizes %\textbf
{the number of model grid cells, number of model grid cells with a high-quality station, and proportion of grid cells with a high-quality station} in each subregion. 

\begin{figure}[!t]
\begin{center}
\includegraphics[trim={0 0 0 0mm}, clip, width =0.9\textwidth]{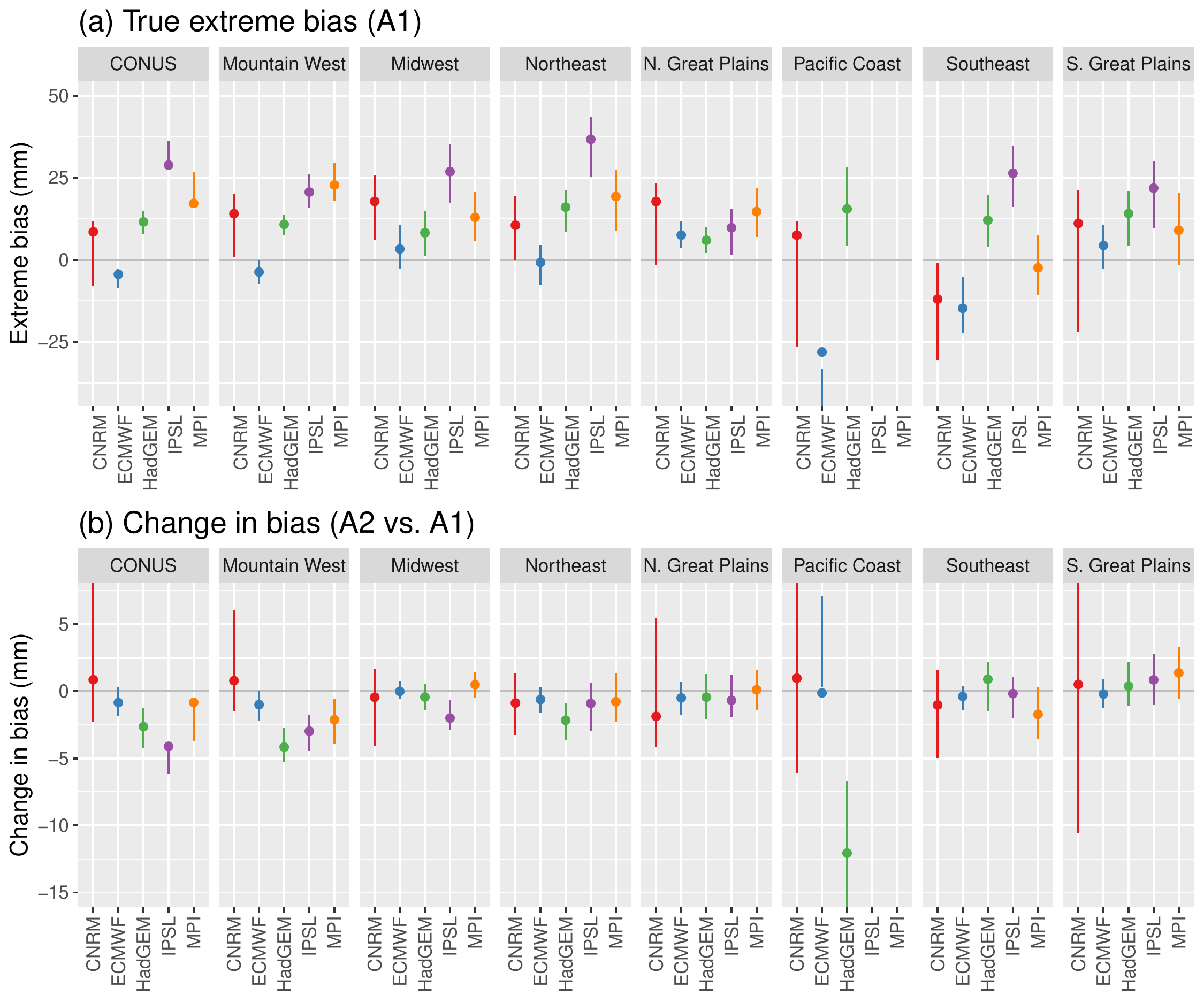}
\caption{True extreme bias in CONUS and the climate subregions for each model (approach A1), with the percent change in extreme bias for approach A2 (ignoring geographic sampling). All estimates show the 95\% basic bootstrap confidence intervals. Note: the extreme bias for IPSL and MPI in the Pacific Coast region is significantly larger than 50mm and the change in bias is significantly less than -20mm (and hence are beyond the limits on the $y$-axis in panels a. and b.).}
\label{meanBias} 
\end{center}
\end{figure}

% \textbf
{First considering the true extreme bias in Figure \ref{meanBias}(a), note that while the models are generally too wet, there are some models and regions that display a dry bias. For example, HadGEM, IPSL, and MPI are almost always too wet (except for MPI in the Southeast); ECMWF is most often too dry (e.g., CONUS, the Southeast, and the Pacific Coast) but sometimes too wet (e.g., the Northern Great Plains). As in Section \ref{section41}, CNRM has very large uncertainties but can be either too wet (e.g., the Midwest) or too dry (the Southeast). Turning to the change in bias due to ignoring geographic sampling, in many cases the change is not significant since the CIs include zero, but when the change \textit{is} significant it is often negative, meaning that ignoring geographic sampling makes the models look drier than they actually are. In other words, the true bias with approach A1 is larger than the bias using approach A2. This is even true for all of CONUS, where the change in bias is significantly nonzero for HadGEM, IPSL, and MPI. The largest biases (in absolute value) and largest changes in bias when accounting for geographic sampling occur in the Pacific Coast, which is not surprising given that it is a highly heterogeneous region in terms of both orography and variability in return values. Interestingly, this is true even though the proportion of the Pacific Coast grid cells with a high-quality station (19\% to 54\% across the model grids) is actually greater than that for the Mountain West (12\% to 43\%; see Table \ref{Ncells}) which has similar degree of orographic variability.
}

\begin{figure}[!t]
\begin{center}
\includegraphics[trim={0 0 0 0mm}, clip, width =0.8\textwidth]{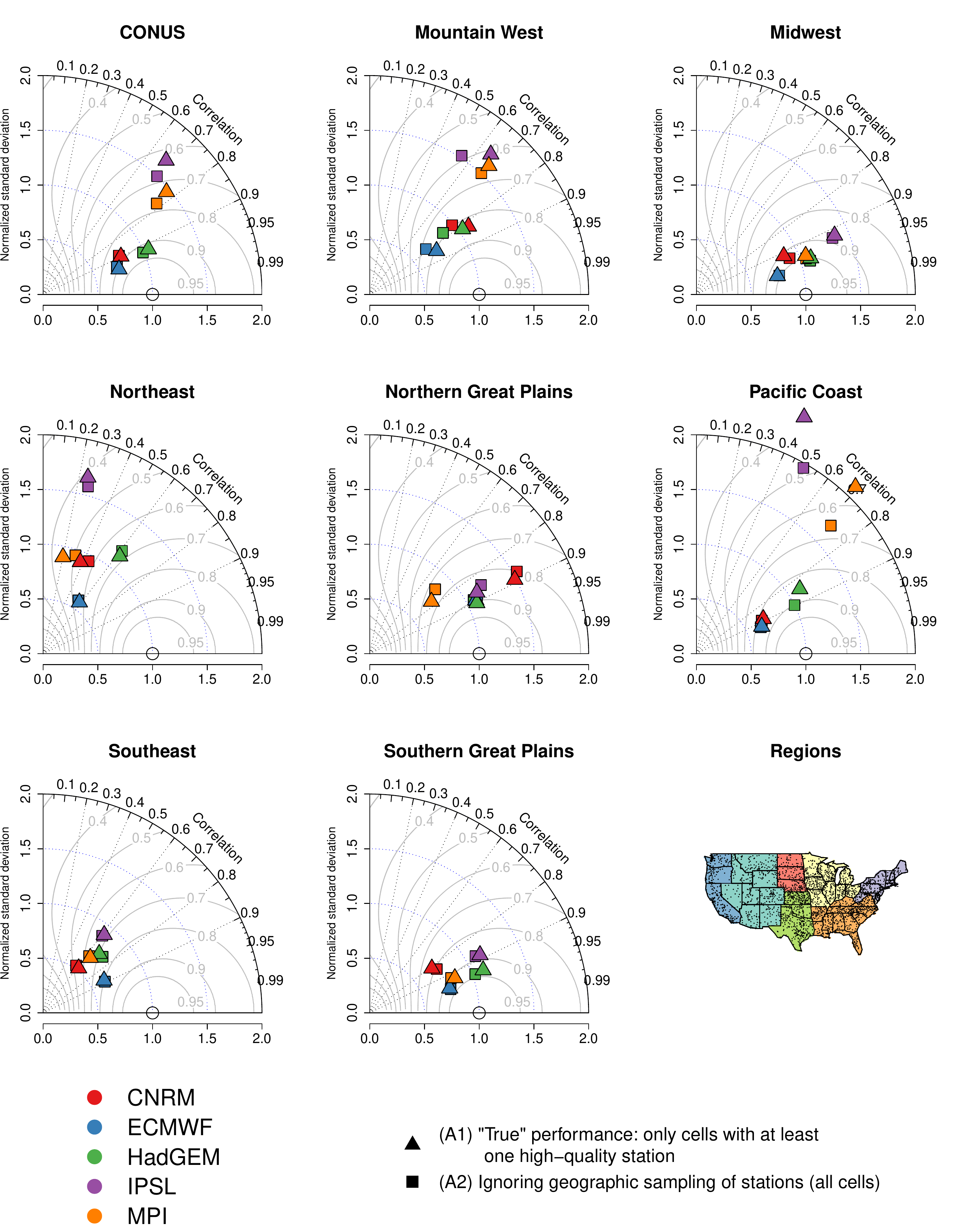}
\caption{Taylor diagrams comparing 20-year return values for CONUS and the climate subregions for the climate models versus regridded L15. The gray curvilinear lines represent the skill score for each model relative to the regridded L15.}
\label{taylorDiags} 
\end{center}
\end{figure}

% \textbf
{Now considering the Taylor Diagrams in Figure \ref{taylorDiags}, the effect of geographic sampling is generally smaller for larger spatial regions, relative to the results for Utah in Figure \ref{taylorDiag_utahkansas}. The effect is particularly small for regions with dense geographic sampling like the Southeast, the Midwest, the Southern Great Plains, and Northern Great Plains. However, the most heterogeneous regions (the Mountain West and Pacific Coast) show the largest effects of sampling, even though as previously mentioned the geographic sampling is relatively good in the Pacific Coast. As with the extreme bias, when it makes a difference the effect of sampling is not the same across the models or these two climate regions. For example, in the Mountain West, there are some models that are more overdispersed (i.e., have too much spatial variability) relative to ignoring sampling (e.g., MPI) while in other cases the true ratio of variability is closer to 1 when accounting for geographic sampling (e.g., HadGEM). The spatial pattern correlation in the Mountain West is generally higher for approach A1, and the true performance of each model generally has a larger skill score relative to A2. In other words, if geographic sampling is ignored, one would conclude that the models are worse than they really are.  Turning to the Pacific Coast region, when geographic sampling has a large effect (for HadGEM, IPSL, and MPI), the true performance of the model generally involves more spatial variability and smaller spatial pattern correlation relative to approach A2. However, in the Pacific Coast, the skill scores are generally higher for A2 than A1: for this region, ignoring geographic sampling leads one to conclude that the skill of the models are better than they actually are.
}

%\textbf
{In summary, as with the case study in Section \ref{section41}, our main point is that the sampling methodology is most important for regions that are highly heterogeneous. Particularly for the extreme bias, the choice of approach A1 vs. A2 can have a large effect regardless of sampling density, but the well-sampled regions generally show little change in the Taylor diagram metrics. Again, the specific effect of ignoring geographic sampling is not systematic, in that the extreme bias, spatial variability, and spatial pattern correlation are impacted in different ways across the various models and climate regions considered.
}

%============================================
\section{Discussion} \label{sec:discussion}

% \begin{itemize}
%     \item Summary of methods and results:
%     \begin{itemize}
%         \item Accounting for geographic sampling does not systematically change the performance of the models (in terms of bias, spatial pattern correlation, and standard deviation) -- any changes appear to be random in nature
%         \item We argue that the sampling should be accounted for in order to yield a more straightforward / direct / appropriate comparison.
%     \end{itemize}
% \textbf
{In this paper, we have highlighted an important issue in comparing extremes from climate model output with observational data, namely that it is important to account for the geographic sampling of weather station data. Our analysis of five-day maxima demonstrates that while accounting for geographic sampling does not systematically change the performance of the models (in terms of bias, spatial pattern correlation, and standard deviation), one should nonetheless account for the sampling in order to yield a more appropriate comparison between models and observations. While our focus in this paper has been on five-day maxima, we expect that similar results would hold for extreme daily or subdaily precipitation and possibly also the mean precipitation climatology. The integrated metrics considered in this paper (namely, extreme bias and Taylor diagrams) are helpful for gaining an overall sense of the models' ability to characterize the extreme climatology, but at the end of the day the quality of the local performance at high resolutions may be much different than what is suggested by, e.g., a spatially-averaged bias. 
}

%\textbf
{The analysis in this paper was motivated by the evaluation of high-resolution climate models, by which we mean models with a $50$km horizontal resolution or finer. Such models are just now becoming widely available, even at the global scale, although current CMIP6-class models generally have a coarser resolution. Accounting for the geographic sampling of weather stations using the criteria outlined in this paper (i.e., only considering grid cells with at least one high-quality station measurement) can have a larger effect as the horizontal resolution increases since many more grid cells will be excluded for a fixed network of high-quality stations. For example, across all of CONUS, the $\sim 20$km HadGEM model has a high quality station in just 22\% of its grid cells, while the coarser $\sim50$km IPSL model has a station in over 60\% of its grid cells. As previously mentioned, accounting for sampling as in this paper would have made no difference at all for a $\sim300$km model like CanESM5 (see Figure \ref{hqGHCNstations}), which has at least one high-quality station in all of its CONUS grid cells. Of course, for lower resolution models like CanESM5 it might be necessary to require a larger number of high-quality stations per grid cell. 
}

% \textbf
{In this paper, we have demonstrated that geographic sampling can make a difference with respect to model evaluation, but we do not suppose it is a perfect solution. Indeed, one of our primary motivations is the idea that a comparison of the climate model versus a gridded product over an area with poor observational sampling could be misleading, since the gridded product does not represent actual measurements of daily precipitation at these locations. However, it must also be admitted that a model quantity at the grid scale is considered a spatial average over sub-grid scales, which offers an equally poor characterization of local values of the variable of interest in topographically complex regions. At the end of the day, a true like-for-like comparison is not possible: regardless of the method used to create a gridded data set, the scales of subgrid parameterizations are not the same as the scale of any station network. Thus, there is always a remaining mismatch. Even when one gets to convection-permitting models, the scale mismatch will remain. Our point in this paper is that accounting for the geographic sampling yields a comparison that is \textit{more} like-for-like, even if mismatches remain.
}

%     \item Talk about extensions to other parts of the globe:
%     \begin{itemize}
%         \item Sampling in places like Kansas is less important
%         \item Cannot say things about flat/homogeneous but very different climatology, e.g., interior of Australia
%     \end{itemize}

% \textbf
{In closing, it is important to note that our definitions of ``well-sampled'' and ``poorly sampled'' in this paper are all relative to the conterminous United States, which is extremely well sampled overall relative to many other land regions. Nonetheless, we have demonstrated the importance of accounting for geographic sampling even for one of the most well-sampled parts of the globe. Sampling considerations will be even more important for the very poorly sampled parts of the world, for example, Africa, South America, northern Asia, and the interior of Australia. The results for Kansas in our case study in Section \ref{section41} bode well for model evaluation in global regions that are poorly sampled but homogeneous, in terms of either the climatology of extreme precipitation or orographic variability. However, it is not immediately obvious how the geographic sampling issue would translate for homogeneous regions that are climatologically very different  from Kansas, for example, desert regions like the interior of Australia or wet regions such as tropical rainforests.
}

\vskip5ex

\noindent\textit{Data availability.}
{The observational data supporting this article are based on publicly available measurements from the National Centers for Environmental Information (\url{ftp://ftp.ncdc.noaa.gov/pub/data/ghcn/daily/} for the GHCN and \url{https://data.nodc.noaa.gov/ncei/archive/data/0129374/daily/} for the L15 product). Model data was obtained for the IPSL and CNRM models from the Earth System Grid (\url{https://esgf-node.llnl.gov/search/cmip6/}) and for the other models by early access to the Jasmin server in the UK. It is expected that all data used in this paper will eventually be uploaded to the CMIP6 data portals by the modeling groups themselves.} %% use this section when having only data sets available

%\codedataavailability{TEXT} %% use this section when having data sets and software code available

%\sampleavailability{TEXT} %% use this section when having geoscientific samples available

%\videosupplement{TEXT} %% use this section when having video supplements available

\vskip2ex

\noindent\textit{Author contribution.}
{Wehner and Risser contributed to developing methodological aspects of the paper. Wehner contributed model data sets. Risser conducted all analyses. Risser and Wehner contributed to write-up of manuscript.} %% this section is mandatory

\vskip2ex

\noindent\textit{Competing interests.} No competing interests are present. %% this section is mandatory even if you declare that no competing interests are present

%\disclaimer{TEXT} %% optional section

\vskip2ex

\noindent\textit{Acknowledgements.}
The authors would like to thank the associate editor and three anonymous reviewers for their comments, which have greatly improved the quality of this manuscript.

This research was supported by the Director, Office of Science, Office of Biological and Environmental Research of the U.S. Department of Energy under Contract No. DE-AC02-05CH11231 and used resources of the National Energy Research Scientific Computing Center (NERSC), also supported by the Office of Science of the U.S. Department of Energy, under Contract No. DE-AC02-05CH11231. 

This document was prepared as an account of work sponsored by the United States Government. While this document is believed to contain correct information, neither the United States Government nor any agency thereof, nor the Regents of the University of California, nor any of their employees, makes any warranty, express or implied, or assumes any legal responsibility for the accuracy, completeness, or usefulness of any information, apparatus, product, or process disclosed, or represents that its use would not infringe privately owned rights. Reference herein to any specific commercial product, process, or service by its trade name, trademark, manufacturer, or otherwise, does not necessarily constitute or imply its endorsement, recommendation, or favoring by the United States Government or any agency thereof, or the Regents of the University of California. The views and opinions of authors expressed herein do not necessarily state or reflect those of the United States Government or any agency thereof or the Regents of the University of California.

 \bibliographystyle{apalike}
 \bibliography{spatial_sampling.bib}

\clearpage
\appendix
\numberwithin{equation}{section}
\numberwithin{table}{section}
\numberwithin{figure}{section}

\section{Supplemental figures}

\begin{figure}[!h]
\begin{center}
\includegraphics[trim={0 0 0 0mm}, clip, width = 0.7\textwidth]{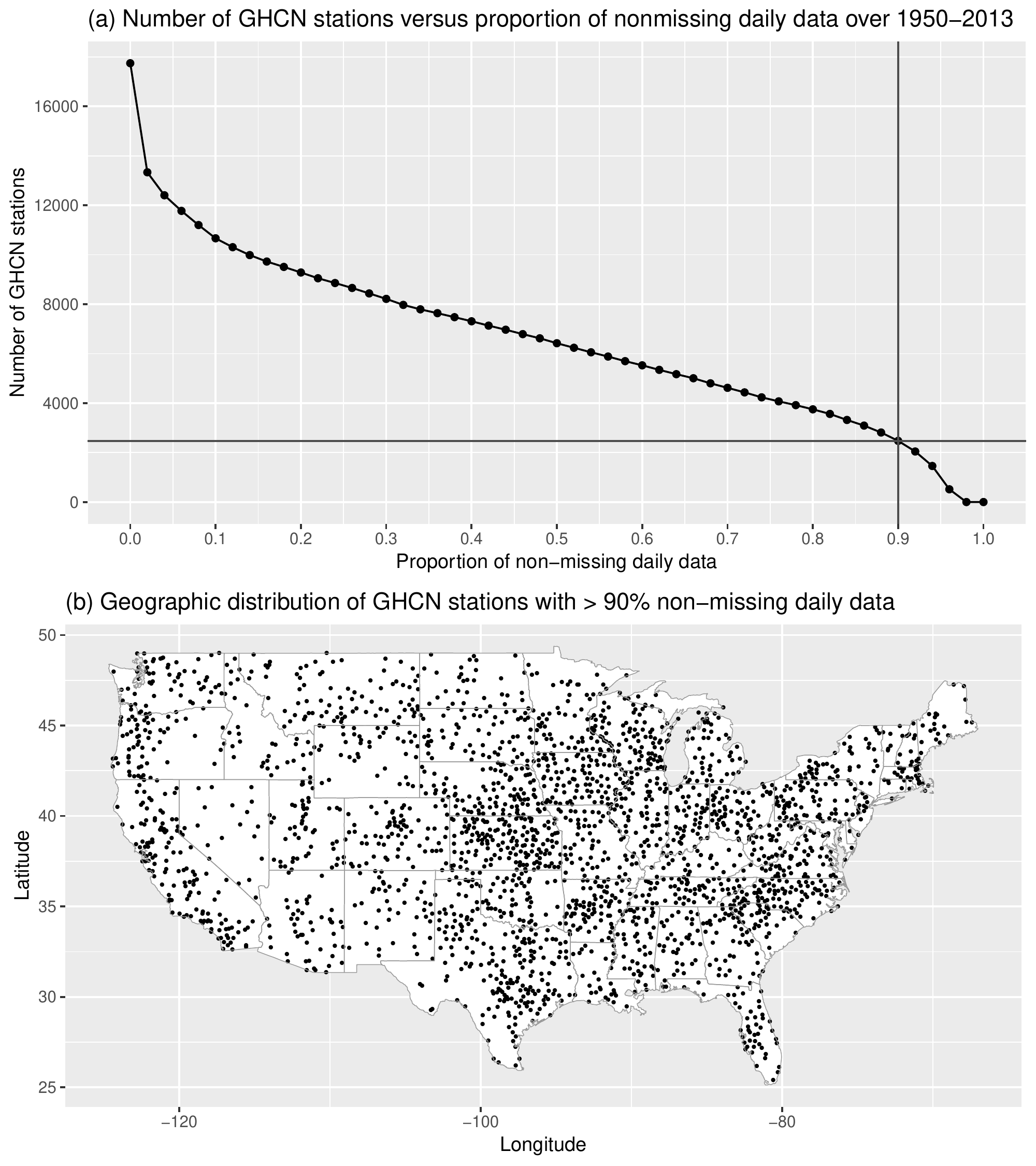}
\caption{The number of GHCN stations versus the proportion of non-missing daily data over 1950-2013 threshold, with the 90\% threshold used in this paper to define the $n = 2474$ ``high-quality'' stations (panel a). The geographic distribution of the high-quality GHCN stations with at least 90\% of non-missing daily measurements over 1950-2013 (panel b).
}
\label{hqGHCNstations}
\end{center}
\end{figure}

\begin{figure}[!h]
\begin{center}
\includegraphics[trim={0 0 0 0mm}, clip, width = 0.9\textwidth]{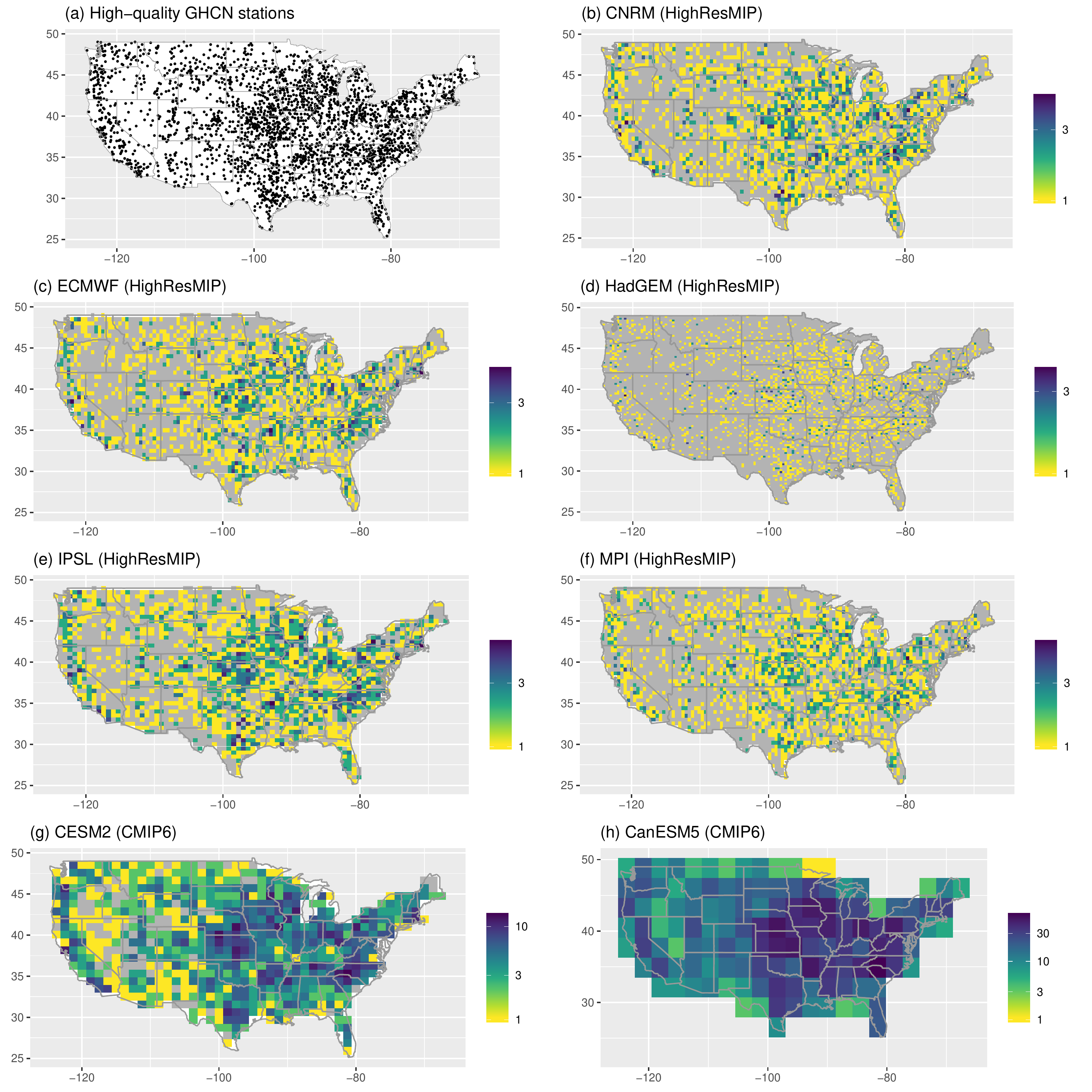}
\caption{The geographic distribution of the high-quality GHCN stations considered in the analysis (panel a.), with the number of high-quality GHCN stations in each model grid cell for the HighResMIP models considered in this paper (panels b.-f.) and two CMIP6 models for comparison (panels g.-h.). Model grid cells without a representative high-quality GHCN station are shown in gray.}
\label{model_masks}
\end{center}
\end{figure}

\begin{figure}[!h]
\begin{center}
\includegraphics[trim={0 0 0 0mm}, clip, width = \textwidth]{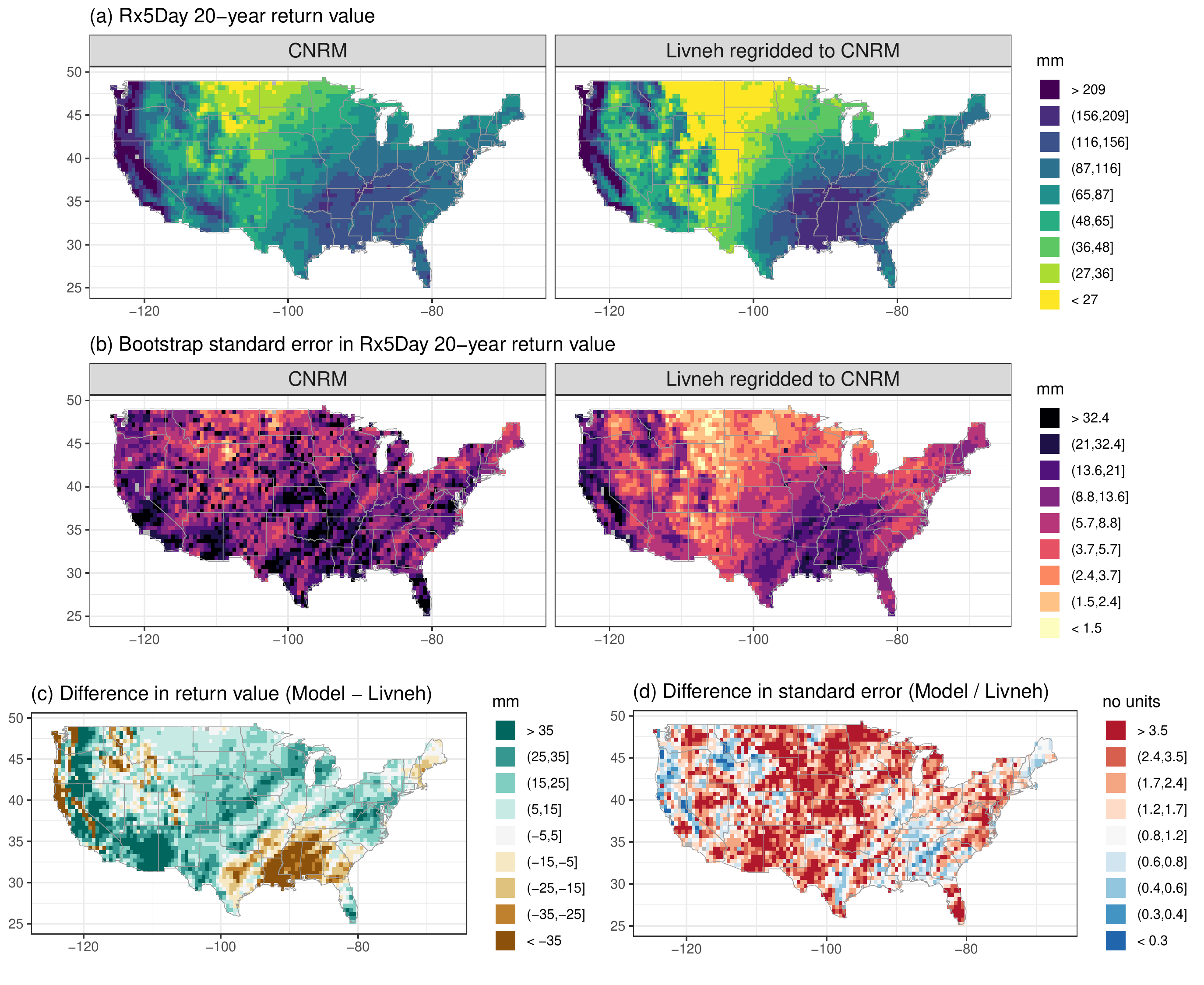}
\caption{Wintertime (DJF) 20-year return values in Rx5Day (mm) for the CNRM model and regridded L15 (panel a), with the bootstrap standard error (mm; panel b). Also shown is the difference in return values (model minus regridded L15) as well as the ratio of standard errors (model divided by regridded L15.}
\label{CNRM}
\end{center}
\end{figure}

\begin{figure}[!h]
\begin{center}
\includegraphics[trim={0 0 0 0mm}, clip, width = \textwidth]{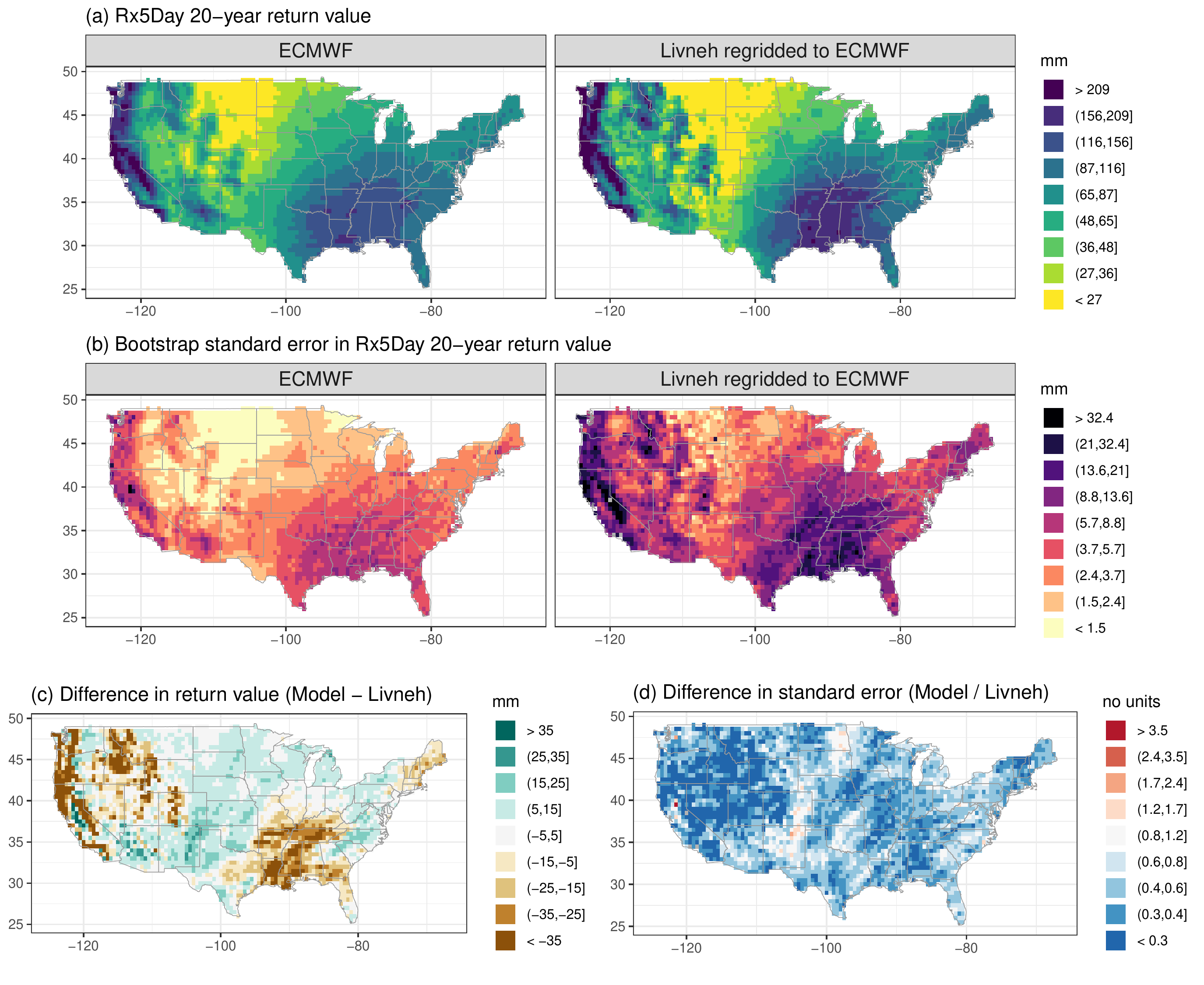}
\caption{Wintertime (DJF) 20-year return values in Rx5Day (mm) for the ECMWF model and regridded L15 (panel a), with the bootstrap standard error (mm; panel b). Also shown is the difference in return values (model minus regridded L15) as well as the ratio of standard errors (model divided by regridded L15.}
\label{ECMWF}
\end{center}
\end{figure}

\begin{figure}[!h]
\begin{center}
\includegraphics[trim={0 0 0 0mm}, clip, width = \textwidth]{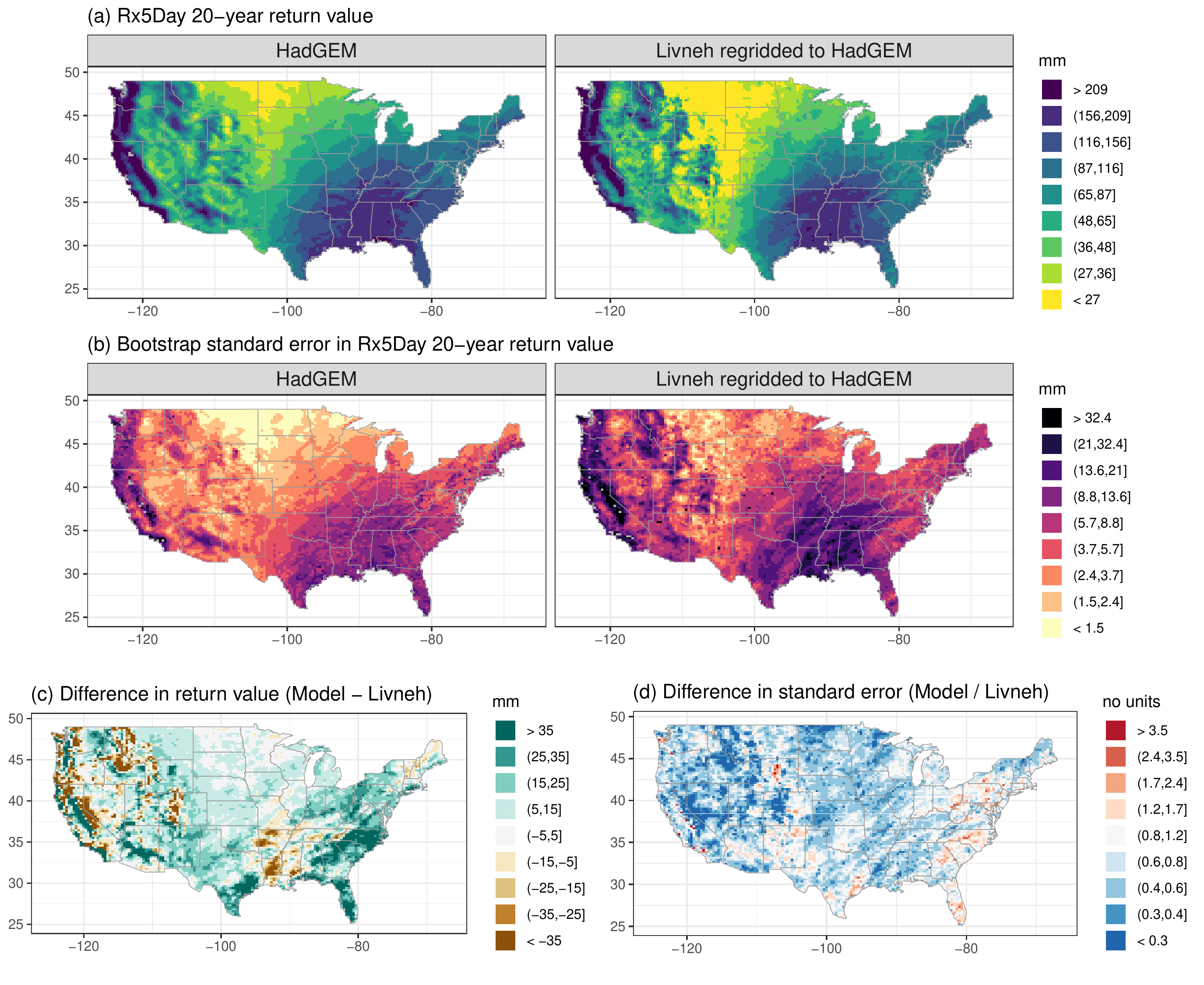}
\caption{Wintertime (DJF) 20-year return values in Rx5Day (mm) for the HadGEM model and regridded L15 (panel a), with the bootstrap standard error (mm; panel b). Also shown is the difference in return values (model minus regridded L15) as well as the ratio of standard errors (model divided by regridded L15.}
\label{HadGEM}
\end{center}
\end{figure}

\begin{figure}[!h]
\begin{center}
\includegraphics[trim={0 0 0 0mm}, clip, width = \textwidth]{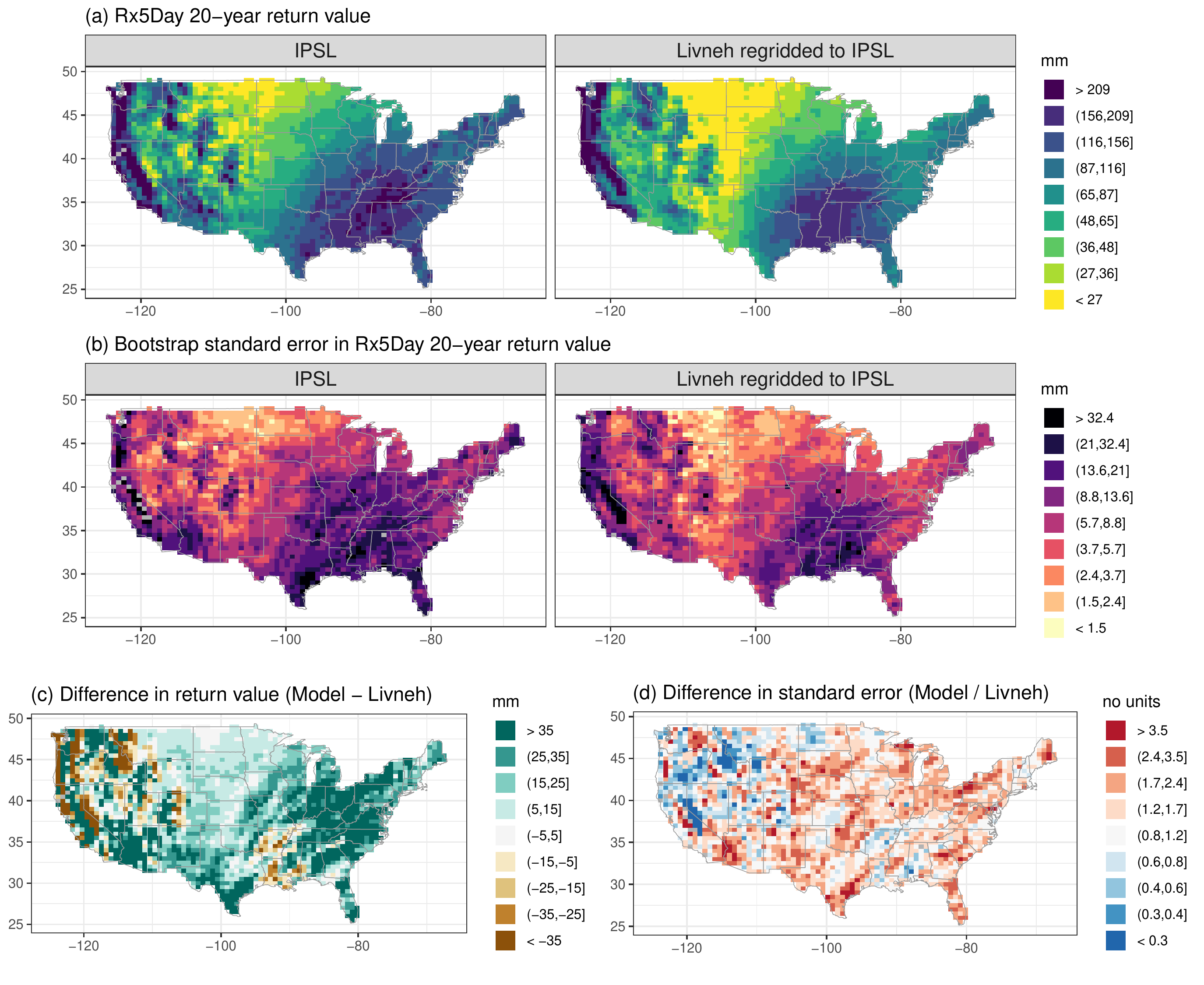}
\caption{Wintertime (DJF) 20-year return values in Rx5Day (mm) for the IPSL model and regridded L15 (panel a), with the bootstrap standard error (mm; panel b). Also shown is the difference in return values (model minus regridded L15) as well as the ratio of standard errors (model divided by regridded L15.}
\label{IPSL}
\end{center}
\end{figure}

\begin{figure}[!h]
\begin{center}
\includegraphics[trim={0 0 0 0mm}, clip, width = \textwidth]{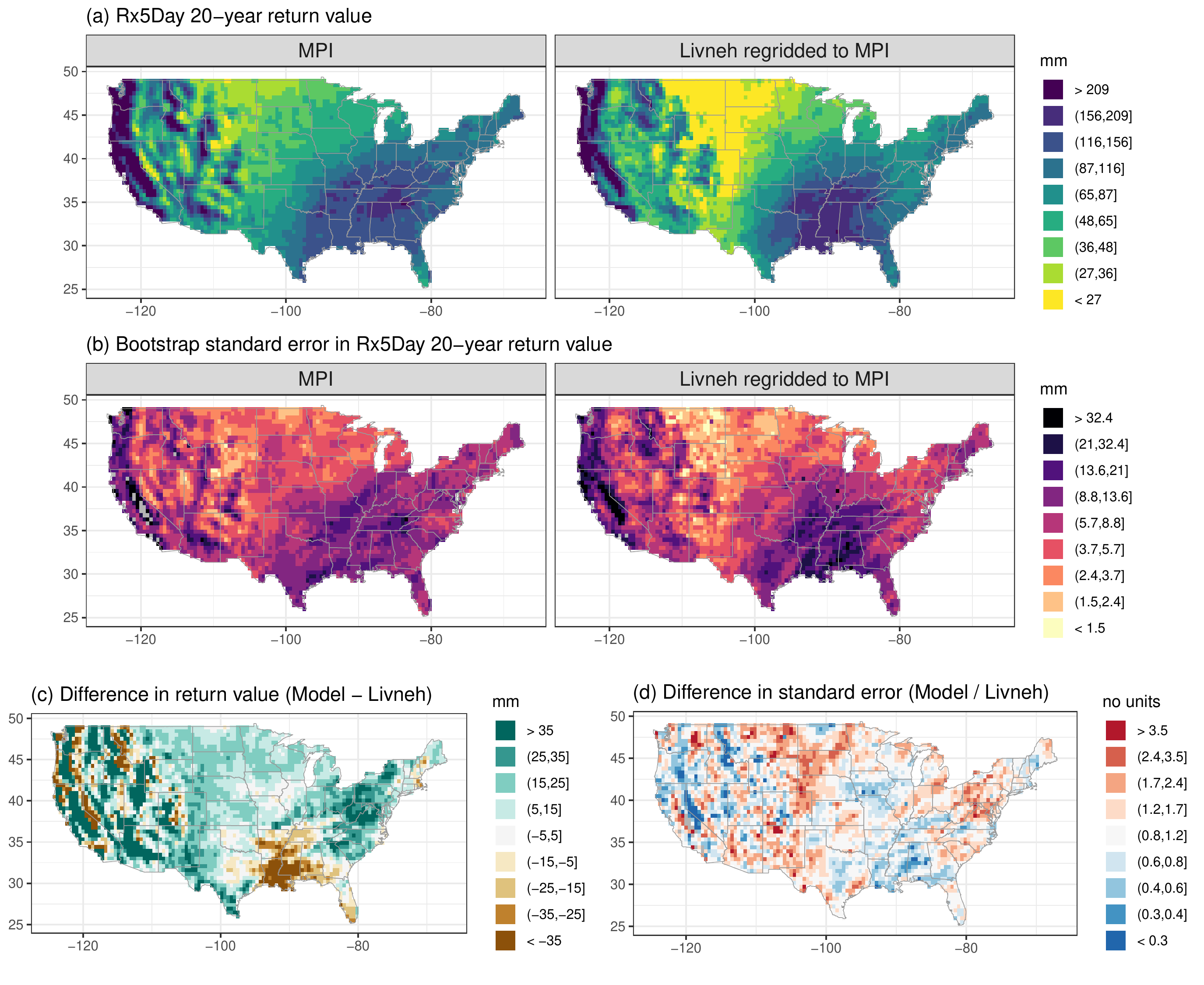}
\caption{Wintertime (DJF) 20-year return values in Rx5Day (mm) for the MPI model and regridded L15 (panel a), with the bootstrap standard error (mm; panel b). Also shown is the difference in return values (model minus regridded L15) as well as the ratio of standard errors (model divided by regridded L15.}
\label{MPI}
\end{center}
\end{figure}

\begin{figure}[!h]
\begin{center}
\includegraphics[trim={0 0 0 0mm}, clip, width = \textwidth]{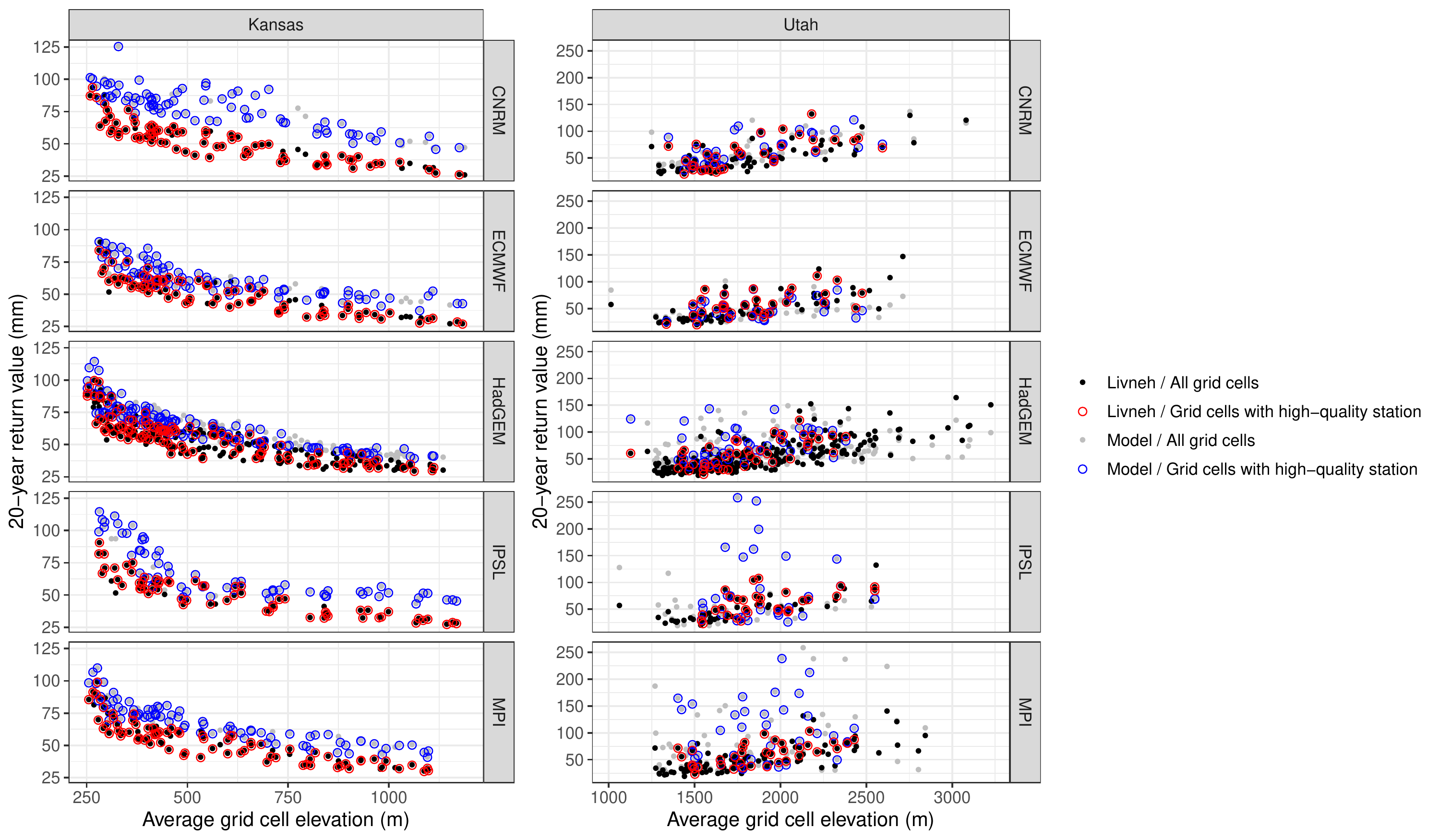}
\caption{The distribution of the grid cell average elevation (in meters; averaged from the GTOPO30 1 km digital elevation product) versus the Rx5Day 20-year return value (in millimeters) for Kansas and Utah across all model grids. Black dots represent the return values for all L15 grid cells, with gray dots representing the climate models. The colored circles (red for L15 and blue for the models) identify the grid cells that have at least one high-quality station and are used in the ``true'' model performance metrics.}
\label{elev_vs_rv}
\end{center}
\end{figure}

\begin{figure}[!h]
\begin{center}
\includegraphics[trim={0 0 0 0mm}, clip, width = \textwidth]{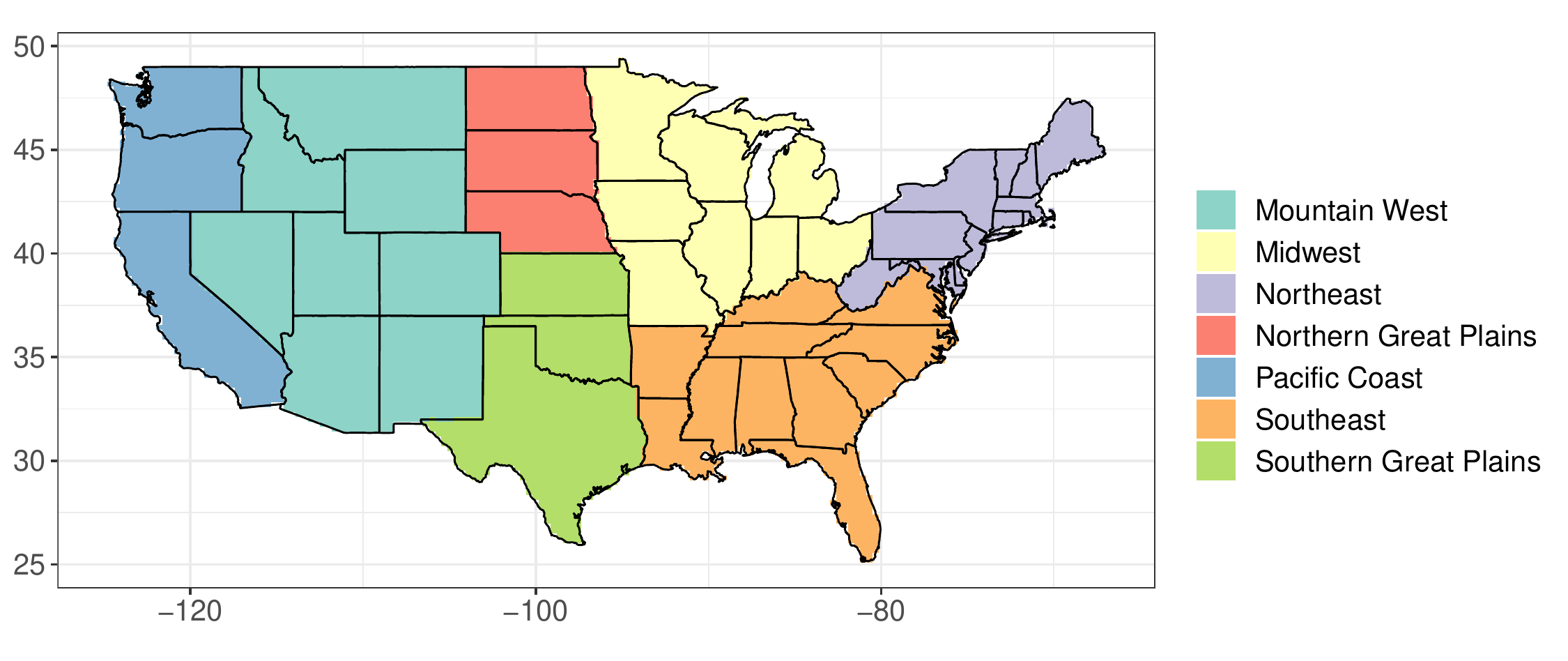}
\caption{The seven subregions used in our analysis. These are a slight variation on the regions defined in the Fourth National Climate Assessment.}
\label{FigA4}
\end{center}
\end{figure}

\begin{table}[t]
\caption{Number of model grid cells in each study region ($N_c$), the number of model grid cells with at least one high-quality GHCN station ($N_{c+s}$), and the proportion of model grid cells with at least one high-quality GHCN station ($P_{c+s}$).}
\begin{center}
\begin{tabular}{|p{0.13\linewidth}||  p{0.065\linewidth}| p{0.065\linewidth}|  p{0.065\linewidth} || p{0.065\linewidth}| p{0.065\linewidth}|  p{0.065\linewidth}|| p{0.065\linewidth}| p{0.065\linewidth}|  p{0.065\linewidth}|} \hline

 &  \multicolumn{3}{c||}{\textbf{Kansas}} & \multicolumn{3}{c||}{\textbf{Utah}} & \multicolumn{3}{c|}{\textbf{CONUS}} \\ \hline
\textbf{Model} & $N_c$ & $N_{c+s}$ & $P_{c+s}$ & $N_c$ & $N_{c+s}$ & $P_{c+s}$ & $N_c$ & $N_{c+s}$ & $P_{c+s}$ \\ \hline \hline
CNRM & 89 & 75 & 0.84 & 92 & 33 & 0.36 & 3256 & 1656 & 0.51\\
\hline
ECMWF & 92 & 76 & 0.83 & 91 & 32 & 0.35 & 3253 & 1660 & 0.51\\
\hline
HadGEM & 268 & 115 & 0.43 & 270 & 42 & 0.16 & 9900 & 2178 & 0.22\\
\hline
IPSL & 68 & 62 & 0.91 & 66 & 30 & 0.45 & 2316 & 1410 & 0.61\\
\hline
MPI & 95 & 78 & 0.82 & 113 & 36 & 0.32 & 3748 & 1729 & 0.46\\
\hline
\multicolumn{10}{c}{ } \\ \hline

 &  \multicolumn{3}{c||}{\textbf{Mountain West}} & \multicolumn{3}{c||}{\textbf{Midwest}} & \multicolumn{3}{c|}{\textbf{Northeast}} \\ \hline
\textbf{Model} & $N_c$ & $N_{c+s}$ & $P_{c+s}$ & $N_c$ & $N_{c+s}$ & $P_{c+s}$ & $N_c$ & $N_{c+s}$ & $P_{c+s}$ \\ \hline \hline
CNRM & 941 & 312 & 0.33 & 513 & 331 & 0.65 & 233 & 136 & 0.58\\
\hline
ECMWF & 945 & 316 & 0.33 & 519 & 341 & 0.66 & 231 & 132 & 0.57\\
\hline
HadGEM & 2887 & 355 & 0.12 & 1559 & 467 & 0.30 & 696 & 183 & 0.26\\
\hline
IPSL & 681 & 292 & 0.43 & 373 & 284 & 0.76 & 161 & 107 & 0.66\\
\hline
MPI & 1113 & 320 & 0.29 & 590 & 351 & 0.59 & 261 & 134 & 0.51\\
\hline

\multicolumn{10}{c}{ } \\ \hline

 &  \multicolumn{3}{c||}{\textbf{Northern Great Plains}} & \multicolumn{3}{c||}{\textbf{Pacific Coast}} & \multicolumn{3}{c|}{\textbf{Southeast}} \\ \hline
\textbf{Model} & $N_c$ & $N_{c+s}$ & $P_{c+s}$ & $N_c$ & $N_{c+s}$ & $P_{c+s}$ & $N_c$ & $N_{c+s}$ & $P_{c+s}$ \\ \hline \hline
CNRM & 274 & 138 & 0.50 & 349 & 152 & 0.44 & 523 & 334 & 0.64\\
\hline
ECMWF & 265 & 128 & 0.48 & 346 & 154 & 0.45 & 519 & 332 & 0.64\\
\hline
HadGEM & 800 & 163 & 0.20 & 1091 & 204 & 0.19 & 1593 & 464 & 0.29\\
\hline
IPSL & 174 & 105 & 0.60 & 254 & 136 & 0.54 & 374 & 280 & 0.75\\
\hline
MPI & 304 & 141 & 0.46 & 411 & 167 & 0.41 & 601 & 357 & 0.59\\
\hline
\multicolumn{10}{c}{ }
\end{tabular}

\begin{tabular}{|p{0.13\linewidth}||  p{0.065\linewidth}| p{0.065\linewidth}|  p{0.065\linewidth} |} \hline

 &  \multicolumn{3}{c|}{\textbf{Southern Great Plains}}  \\ \hline
\textbf{Model} & $N_c$ & $N_{c+s}$ & $P_{c+s}$  \\ \hline \hline
CNRM & 423 & 253 & 0.60\\
\hline
ECMWF & 428 & 257 & 0.60\\
\hline
HadGEM & 1274 & 342 & 0.27\\
\hline
IPSL & 299 & 206 & 0.69\\
\hline
MPI & 468 & 259 & 0.55\\
\hline
\end{tabular}
\label{Ncells}
\end{center}
\end{table}

\end{document}